\documentclass{aa}  

\usepackage{graphicx}
\usepackage{txfonts}
\usepackage{hyperref}
\pdfoutput=1
\usepackage{amsfonts}
\usepackage{amsmath}
\usepackage{amssymb}
\usepackage{xcolor}
\usepackage{enumitem}
\usepackage{graphicx}
\usepackage{gensymb}
\usepackage{fancyhdr}
\usepackage{times}
\usepackage{ulem}
\usepackage{subcaption} 
\usepackage{float} 
\usepackage{color}
\usepackage{booktabs}
\usepackage{lscape}
\usepackage{xspace}

\definecolor{purple}{RGB}{76, 0,153}

\newcommand{\dd}{{\rm d}}
\newcommand{\diff}{{\rm d}}

\newcommand{\br}[1]{\left( #1 \right)}
\newcommand{\bc}[1]{\left\{ #1 \right\}}
\newcommand{\eqa}[1]{\begin{align}   #1 \end{align}}

\newcommand{\bb}[1]{\left[ #1 \right]}
\newcommand{\vek}[1]{\mbox{\boldmath $#1$}}

\newcommand{\software}[1]{\texttt{#1}}
\newcommand{\tttp}{$3\times2$pt\xspace}
\newcommand{\kpoff}{\ensuremath{3.1\,\sigma}\xspace}
\newcommand{\kpoffperc}{\ensuremath{8.3 \pm 2.6\%}\xspace}
\newcommand{\kSeightval}{\ensuremath{0.766^{+0.020}_{-0.014}}\xspace}
\newcommand{\kOmegamval}{\ensuremath{0.305^{+0.010}_{-0.015}}\xspace}
\newcommand{\ksigmaeightval}{\ensuremath{0.76^{+0.025}_{-0.020}}\xspace}

\newcommand{\klogR}{\ensuremath{3.1\pm0.3}\xspace}
\newcommand{\kR}{\ensuremath{23\pm6}\xspace}
\newcommand{\klogS}{\ensuremath{-2.0\pm0.1}\xspace}
\newcommand{\klogSPTE}{\ensuremath{0.08\pm0.02}\xspace}
\newcommand{\klogSPTEsigma}{\ensuremath{1.8\pm0.1\,\sigma}\xspace}

\newcommand{\LCDM}{$\Lambda$CDM\xspace}

\newcommand{\be}{\begin{equation}}  \newcommand{\ee}{\end{equation}}

\usepackage{environ}
\NewEnviron{splitequation}{%
    \begin{equation}\begin{split}
        \BODY
    \end{split}\end{equation}}
    
\begin{document} 

   \title{KiDS-1000 Cosmology: Multi-probe weak gravitational lensing and spectroscopic galaxy clustering constraints}

   \author{Catherine Heymans \inst{1,2}\thanks{Catherine Heymans: heymans@roe.ac.uk} 
   \and Tilman Tr\"oster\inst{1}\thanks{Tilman Tr\"oster: ttr@roe.ac.uk} 
   \and Marika Asgari\inst{1} 
   \and Chris Blake\inst{3}
   \and Hendrik Hildebrandt\inst{2}
   \and Benjamin Joachimi\inst{4}
   \and Konrad Kuijken\inst{5}
   \and Chieh-An Lin\inst{1}
   \and Ariel~G.~S\'anchez\inst{6}
   \and Jan Luca van den Busch\inst{2}
   \and Angus~H.~Wright\inst{2}
   \and Alexandra Amon\inst{7}
   \and Maciej Bilicki\inst{8}
   \and Jelte de Jong\inst{9}
   \and Martin Crocce\inst{10,11}
   \and Andrej Dvornik\inst{2}
   \and Thomas Erben\inst{12}
   \and Maria Cristina Fortuna\inst{5}
   \and Fedor Getman\inst{13}
   \and Benjamin Giblin\inst{1}
   \and Karl Glazebrook\inst{3}
   \and Henk Hoekstra\inst{5}
   \and Shahab Joudaki\inst{14}
   \and Arun Kannawadi\inst{15,5}
   \and Fabian K\"ohlinger\inst{2}
   \and Chris Lidman\inst{16}	
   \and Lance Miller\inst{14}
   \and Nicola~R.~Napolitano\inst{17}
   \and David Parkinson\inst{18}
   \and Peter Schneider \inst{12}
   \and HuanYuan Shan\inst{19,20}
   \and Edwin A. Valentijn\inst{9}
   \and Gijs Verdoes Kleijn\inst{9}
   \and Christian Wolf \inst{16}
          }
\institute{Institute for Astronomy, University of Edinburgh, Royal Observatory, Blackford Hill, Edinburgh, EH9 3HJ, UK 
   \and
   Ruhr-Universit{\"a}t Bochum, Astronomisches Institut, German Centre for Cosmological Lensing (GCCL), Universit{\"a}tsstr.  150, 44801, Bochum, Germany
   \and 
   Centre for Astrophysics \& Supercomputing, Swinburne University of Technology, P.O. Box 218, Hawthorn, VIC 3122, Australia
   \and
   Department of Physics and Astronomy, University College London, Gower Street, London WC1E 6BT, UK
   \and
   Leiden Observatory, Leiden University, Niels Bohrweg 2, 2333 CA Leiden, the Netherlands
   \and
   Max-Planck-Institut f\"ur extraterrestrische Physik, Postfach 1312, Giessenbachstrasse 1, D-85741 Garching, Germany
   \and
   Kavli Institute for Particle Astrophysics \& Cosmology, P. O. Box 2450, Stanford University, Stanford, CA 94305, USA
   \and 
   Center for Theoretical Physics, Polish Academy of Sciences, al. Lotnik\'{o}w 32/46, 02-668, Warsaw, Poland
   \and 
   Kapteyn Astronomical Institute, University of Groningen, PO Box 800, 9700 AV Groningen, the Netherlands
   \and
   Institute of Space Sciences (ICE, CSIC), Campus UAB, Carrer de Can Magrans, s/n,  E-08193 Barcelona, Spain 
   \and
   Institut d'Estudis Espacials de Catalunya (IEEC),  Carrer Gran Capita 2, E-08034, Barcelona, Spain
   \and
   Argelander-Institut f\"ur Astronomie, Universit\"at Bonn, Auf dem H\"ugel 71, D-53121 Bonn, Germany
   \and
   INAF - Astronomical Observatory of Capodimonte, Via Moiariello 16, 80131 Napoli, Italy
   \and
   Department of Physics, University of Oxford, Denys Wilkinson Building, Keble Road, Oxford OX1 3RH, UK
    \and
   Department of Astrophysical Sciences, Princeton University, 4 Ivy Lane, Princeton, NJ 08544, USA
    \and
   Research School of Astronomy and Astrophysics, Australian National University, Canberra ACT 2600, Australia
   \and
   University of Chinese Academy of Sciences, Beijing 100049, China
   \and
   Korea Astronomy and Space Science Institute, 776 Daedeokdae-ro, Yuseong-gu, Daejeon 34055, Republic of Korea
   \and
   Shanghai Astronomical Observatory (SHAO), Nandan Road 80, Shanghai 200030, China
   \and
   University of Chinese Academy of Sciences, Beijing 100049, China
   }

 
  \abstract{We present a joint cosmological analysis of weak gravitational lensing observations from the Kilo-Degree Survey (KiDS-1000), with redshift-space galaxy clustering observations from the Baryon Oscillation Spectroscopic Survey (BOSS) and galaxy-galaxy lensing observations from the overlap between KiDS-1000, BOSS, and the spectroscopic 2-degree Field Lensing Survey (2dFLenS).  This combination of large-scale structure probes breaks the degeneracies between cosmological parameters for individual observables, resulting in a constraint on the structure growth parameter $S_8=\sigma_8 \sqrt{\Omega_{\rm m}/0.3} =$ \kSeightval, which has the same overall precision as that reported by the full-sky cosmic microwave background observations from {\it Planck}.   The recovered $S_8$ amplitude is low, however, by \kpoffperc relative to {\it Planck}.   This result builds from a series of KiDS-1000 analyses where we validate our methodology with variable depth mock galaxy surveys,  our lensing calibration with image simulations and null-tests, and our optical-to-near-infrared redshift calibration with multi-band mock catalogues and a spectroscopic-photometric clustering analysis.   The systematic uncertainties identified by these analyses are folded through as nuisance parameters in our cosmological analysis.  Inspecting the offset between the marginalised posterior distributions, we find that the $S_8$-difference with {\it Planck} is driven by a tension in the matter fluctuation amplitude parameter, $\sigma_8$.   
We quantify the level of agreement between the cosmic microwave background and our large-scale structure constraints using a series of different metrics, finding differences with a significance ranging between $\sim\! 3\,\sigma$, when considering the offset in $S_{8}$, and $\sim\! 2\,\sigma$, when  considering the full multi-dimensional parameter space. }

 \keywords{gravitational lensing: weak, methods: data analysis, methods: statistical, surveys, cosmology: observations}

   \titlerunning{KiDS-1000: 3x2pt}
   \authorrunning{Heymans, Tr\"oster \& the KiDS Collaboration et al.}
   \maketitle
%
\section{Introduction}
\label{sec:intro}

Observations of the cosmic microwave background (CMB) have delivered high-precision
constraints for the cosmological parameters of the flat, cold dark
matter, and cosmological constant model of the Universe
\citep[$\Lambda$CDM,][]{planck/etal:2018}.  With only six free
parameters, this flat $\Lambda$CDM model provides an exquisite fit to observations of
the anisotropies in the CMB.    The same model predicts a range of
different observables in the present-day Universe, including the cosmic expansion rate \citep{weinberg/1972}, and the
distribution of, and gravitational lensing by, large-scale
structures \citep{peebles/1980,bartelmann/schneider:2001,eisenstein/etal:2005}.  
In most cases there
is agreement between the measured cosmological parameters of the flat $\Lambda$CDM model, when comparing those
constrained at the CMB epoch with those constrained through a variety of
lower-redshift probes \citep[see the discussion in][and references
therein]{planck/etal:2018}.   Recent improvements in the statistical
precision of the lower-redshift probes have,
however, revealed some statistically significant differences.  Most
notably, a $4.4\,\sigma$ difference in the value of the Hubble constant,
$H_0$, has been reported using distance ladder estimates in \citet{riess/etal:2019}.  If this difference cannot be
attributed to systematic errors in either experiment, or both, this result
suggests that the flat $\Lambda$CDM model is incomplete.

Many
extensions have been proposed to reconcile the observed differences between
high- and low-redshift probes \citep[see for
example][]{riess/etal:2016,poulin/etal:2018,divalentino/etal:2020}.  All, however, require
additional components
to the cosmological model that move it even further away from the
standard model of particle physics, a model that already struggles to motivate
the existence of cold dark matter and a cosmological constant.  
As the statistical power of the observations continues to improve, focus has shifted to establishing
a full understanding of all systematic errors and to the development of mitigation approaches, 
in preparation for the high-precision `full-sky' imaging and spectroscopic
cosmology surveys of the 2020s \citep[{\it Euclid},][]{laureijs/etal:2011,lsst/etal:2009,DESI/etal:2016}.

We present a multi-probe `same-sky' analysis of the
evolution of large-scale structures, using overlapping spectroscopic and optical-to-near-infrared imaging surveys.
Our first observable is the weak gravitational lensing of background
galaxies by foreground large-scale
structures, 
known as `cosmic shear'.    Our second
observable is the anisotropic clustering of galaxies within these
large-scale structures, combining measurements of both redshift-space
distortions and baryon acoustic
oscillations.   Our third observable is the weak gravitational lensing of background
galaxies by the matter surrounding foreground galaxies, known as
`galaxy-galaxy lensing'.   As these three sets of two-point
statistics are analysed simultaneously, this combination of probes
is usually referred to as a `\tttp' analysis. 

Each observable in our multi-probe analysis is subject to systematic
uncertainties.  For a cosmic shear analysis, the observable is a
combination of the true cosmological signal with a low-level signal
arising from the intrinsic alignment of galaxies, as well as potential residual
correlations in the data induced by the atmosphere, telescope, and
camera.   The signal can also be scaled by both shear
 and photometric redshift measurement calibration errors
 \citep[see][and references therein]{mandelbaum:2018}.   For a galaxy
   clustering analysis, the observable is the true
   cosmological signal modulated by an uncertain galaxy bias function.  
   This function maps how
   the galaxies trace the
   underlying total matter distribution \citep[see][and references
   therein]{desjacques/etal:2018}.  It can be non-linear and evolves with redshift. 
   The cosmological clustering
   also needs to be accurately distinguished from artificial clustering in the galaxy sample,
   arising from potentially uncharacterised inhomogeneities in the target selection \citep[see for example][]{ross/etal:2012}. 
   Finally, the galaxy-galaxy
   lensing analysis is subject to the systematics that impact both the
   cosmic shear and clustering analyses.

   When analysing these
   observables in combination
   the different astrophysical and systematic dependencies allow for some degree of
   self-calibration \citep{bernstein/jain:2004, hu/jain:2004,
     bernstein:2009,joachimi/bridle:2010}.  Adopting `same-sky'
   surveys, in which imaging for weak lensing observables overlaps with
   spectroscopy for anisotropic galaxy clustering measurements,
   also allows for their cross-correlation.  Such a survey design therefore
   presents a robust
   cosmological tool that can calibrate and mitigate systematic and astrophysical
   uncertainties through a series of nuisance parameters.   In
   addition to enhanced control over systematics, this combination of probes
   breaks cosmological parameter degeneracies from each individual
   probe. 
   For a flat $\Lambda$CDM model,
   this leads to significantly tighter constraints on the matter fluctuation amplitude 
   parameter, $\sigma_8$, and the matter density parameter, $\Omega_{\rm m}$, whilst also decreasing the
   uncertainty on the recovered dark energy equation of state
   parameter in extended cosmology scenarios \citep{hu/jain:2004,gaztanaga/etal:2012}.
   
   Three variants of a joint `\tttp' analysis have been
   conducted to date.  \citet{vanuitert/etal:2018} present a joint power-spectrum
   analysis of the Kilo-Degree Survey \citep[KiDS,][]{kuijken/etal:2015} with
   the Galaxy And Mass Assembly survey
   \citep[GAMA,][]{liske/etal:2015}, incorporating projected
   angular clustering measurements.   \citet{joudaki/etal:2018}
   present a joint analysis of KiDS with the
   2-degree Field Lensing Survey \citep[2dFLenS,][]{blake/etal:2016}
   and the overlapping area in the Baryon Oscillation Spectroscopic Survey \citep[BOSS,][]{alam/etal:2015}, incorporating
   redshift-space clustering measurements.  \citet{abbott/etal:2018}
   present a joint real-space lensing-clustering analysis of the Dark
   Energy Survey \citep[DES Y1,][]{drlicawagner/etal:2018}, using a high-quality
   photometric redshift sample of luminous red galaxies for their projected
   angular clustering measurements.  
   In all three cases a
   linear galaxy bias model was adopted. 
   
   In this analysis we enhance and build upon the advances of previous `\tttp' studies.   We analyse the most recent KiDS data release \citep[KiDS-1000,][]{kuijken/etal:2019}, more than doubling the
   survey area from previous KiDS studies.   We utilise the full BOSS
   area and the `full-shape' anisotropic clustering measurements of \citet{sanchez/etal:2017},
   incorporating information from both redshift-space distortions
   and the baryon acoustic oscillation as our galaxy clustering probe.   We adopt a non-linear
   evolving galaxy bias model, derived from renormalised perturbation theory
   \citep{crocce/scoccimarro:2006, chan/etal:2012}.  
   We maximise the signal-to-noise in our 
   KiDS-BOSS galaxy-galaxy lensing analysis, 
   by including additional overlapping spectroscopy of BOSS-like galaxies from 2dFLenS.

This paper is part of the KiDS-1000 series.  The KiDS-1000 photometry and imaging is presented in \citet{kuijken/etal:2019}.  The core weak lensing data products are presented and validated in \citet[shear measurements,][]{giblin/etal:inprep},  and  \citet[redshift measurements,][]{hildebrandt/etal:inprep}.   \citet{asgari/etal:inprep} conduct the cosmic shear analysis using a range of different two-point statistics, and \citet{joachimi/etal:inprep} detail the methodology behind our `\tttp'  
   analysis, with a particular focus on pipeline validation and accurate covariance matrices.   In this analysis we constrain the cosmological parameters of the flat $\Lambda$CDM model.   A range of different extensions to the $\Lambda$CDM model are considered in \citet{troester/etal:inprep}, including varying dark energy, neutrino mass, spatial curvature and various modified gravity scenarios \citep{bose/etal:2020}.
   
In this paper we review the data and provide a concise summary of the findings of the KiDS-1000 series of papers in Section~\ref{sec:data}.   
 We present our joint cosmological constraints in Section~\ref{sec:results} and conclude in Section~\ref{sec:conc}.  Appendices tabulate the galaxy properties (\ref{app:properties}), the adopted cosmological parameter priors (\ref{app:priors}),  and the cosmological parameter constraints (\ref{app:parameter-constraints}).   They also discuss: the choice of intrinsic galaxy alignment model (\ref{app:IAmodel}); a series of sensitivity tests (\ref{app:sensitivity}); the expected differences between parameter constraints for overlapping weak lensing surveys (\ref{app:expectedoffsets}); a range of different `tension' metrics (\ref{app:tensionest}); the redundancy, validation and software review for our pipeline (\ref{app:codereview}); and the minor analysis additions that were included after the analysis was formally unblinded (\ref{app:unblinding}).

\section{Data and methodology}
\label{sec:data}

\subsection{Surveys:  KiDS, BOSS, and 2dFLenS}
\label{sec:surveys}

The Kilo-Degree Survey \citep[KiDS,][]{dejong/etal:2013}, covers $1350\,\mathrm{deg}^{2}$ split into two fields, one
equatorial and one southern.    Matched-depth imaging in nine bands spans the optical,
$ugri$, through to the near-infrared, $ZYJHK_{\rm s}$, where the
near-infrared imaging was taken as part of the KiDS partner survey
VIKING \citep[the VISTA Kilo-degree INfrared Galaxy
survey,][]{edge/etal:2013}.  High-quality seeing was
routinely allocated to the primary KiDS $r$-band VST-OmegaCAM observations, resulting in a
mean $r$-band seeing of 0.7 arcseconds, with a time-allocated maximum of 0.8
arcseconds.  This combination of full-area spatial and wavelength
resolution over a thousand square degrees
provides a unique weak lensing survey that allows for enhanced
control of systematic errors \citep{giblin/etal:inprep, hildebrandt/etal:inprep}.
This analysis uses data from the fourth KiDS
data release of $1006\,\mathrm{deg}^{2}$ of imaging, (hence the name KiDS-1000), which has an effective
area, after masking, of $777\,\mathrm{deg}^{2}$.  KiDS is a public survey from the European Southern
Observatory, with data products freely accessible through the ESO
archive\footnote{KiDS-DR4 data access: \href{http://kids.strw.leidenuniv.nl/DR4}{kids.strw.leidenuniv.nl/DR4}}.   

\citet{giblin/etal:inprep} present a series of null-tests to validate the KiDS-1000 shear catalogue\footnote{KiDS-1000 Shear Catalogue: \href{http://kids.strw.leidenuniv.nl/DR4/lensing.php}{kids.strw.leidenuniv.nl/DR4/lensing.php}} in five tomographic bins spanning a 
photometric redshift range of $0.1 < z_{\rm B} \leq 1.2$ (see Appendix~\ref{app:properties} for details of the properties of each bin).  
Meeting their requirement that any systematic detected induces less than a $0.1\,\sigma$ change in the inferred 
cosmic shear constraints on the clustering cosmological parameter $S_8 = \sigma_8\sqrt{\Omega_{\rm m}/0.3}$, they conclude that the shear catalogue is `science-ready', with no significant non-lensing B-mode distortions detected.  \citet{kuijken/etal:2015} present the catalogue-level blinding methodology that we adopted to introduce $\pm 2\sigma$ differences in the recovered value of $S_8$ in order
to retain team ignorance over the final cosmological results until all 
analysis decisions were finalised (for further details on blinding see Appendix~\ref{app:unblinding}).
\citet{hildebrandt/etal:inprep} present the KiDS-1000 photometric redshift calibration.  This is determined using the self-organising map (SOM) methodology of \citet{wright/etal:2020}, and is validated with a cross-correlation clustering analysis, following \citet{vandenbusch/etal:2020}.
The SOM identifies and excludes any galaxies that are poorly represented in the spectroscopic calibration sample, in terms of their nine-band colours and magnitudes. The resulting `gold' photometric sample, with an accurately calibrated redshift distribution, is then re-simulated in the KiDS image simulations of \citet{kannawadi/etal:2019} in order to determine the shear calibration corrections for each tomographic bin, and an associated uncertainty \citep[see][for full details]{giblin/etal:inprep,hildebrandt/etal:inprep}.
 
The Baryon Oscillation Spectroscopic Survey
\citep[BOSS,][]{alam/etal:2015}, spans an effective area of $9329\,\mathrm{deg}^{2}$, with spectroscopic redshifts for 1.2 million luminous red
galaxies (LRG) in the redshift range $0.2<z<0.9$.   A range of
different statistical analyses of the clustering of BOSS galaxies have been used in combination with CMB
measurements, to set tight constraints on extensions to the standard
flat $\Lambda$CDM model \citep[see][and references
therein]{alam/etal:2017,eBOSS/etal:2020}.   We adopt the anisotropic clustering
measurements of \citet{sanchez/etal:2017} in this multi-probe analysis.
BOSS only overlaps with the equatorial stripe
of the KiDS survey, with $409\,\mathrm{deg}^{2}$ of the BOSS survey lying within
the KiDS-1000 footprint.  BOSS galaxies in this overlapping region are used as lenses in
our galaxy-galaxy lensing analysis, with an effective lens number density of $0.031\,\mathrm{arcmin}^{-2}$ (see Appendix~\ref{app:properties} for details).  BOSS is a public survey from the third Sloan
Digital Sky Survey \citep{york/etal:2000}, and we analyse data from the twelfth data release\footnote{BOSS data access: \href{https://data.sdss.org/sas/dr12/boss/lss/}{data.sdss.org/sas/dr12/boss/lss/}} \citep[DR12,][]{alam/etal:2015}. 

The 2-degree Field Lensing Survey
\citep[2dFLenS,][]{blake/etal:2016}, spans $731\,\mathrm{deg}^{2}$, with
spectroscopic redshifts for $70\,000$ galaxies out to $z<0.9$.   This
galaxy redshift survey from the Anglo-Australian Telescope (AAT) was designed
to target areas already mapped by weak lensing surveys to facilitate `same-sky'
lensing-clustering analyses
\citep{johnson/etal:2017,amon/etal:2018,joudaki/etal:2018, blake/etal:2020}.
We use data from the 2dFLenS LRG sample that was targeted to match
the BOSS-LRG selection, but with sparser sampling.  2dFLenS
thus provides an additional sample of BOSS-like galaxies in the KiDS
southern stripe where there is $425\,\mathrm{deg}^{2}$ of overlap within
the KiDS-1000 footprint.  2dFLenS galaxies in this overlapping region are used as lenses in
our galaxy-galaxy lensing analysis, with an effective lens number density of $0.012\,\mathrm{arcmin}^{-2}$ (see Appendix~\ref{app:properties} for details).  2dFLenS was an AAT Large Programme that has been made public\footnote{2dFLenS data
  access: \href{http://2dflens.swin.edu.au/data.html}{2dflens.swin.edu.au/data.html}}.   

\subsection{Cosmic shear}
\label{sec:cosmic_shear}

\begin{figure*}
\begin{centering}
        \includegraphics[width=\textwidth]{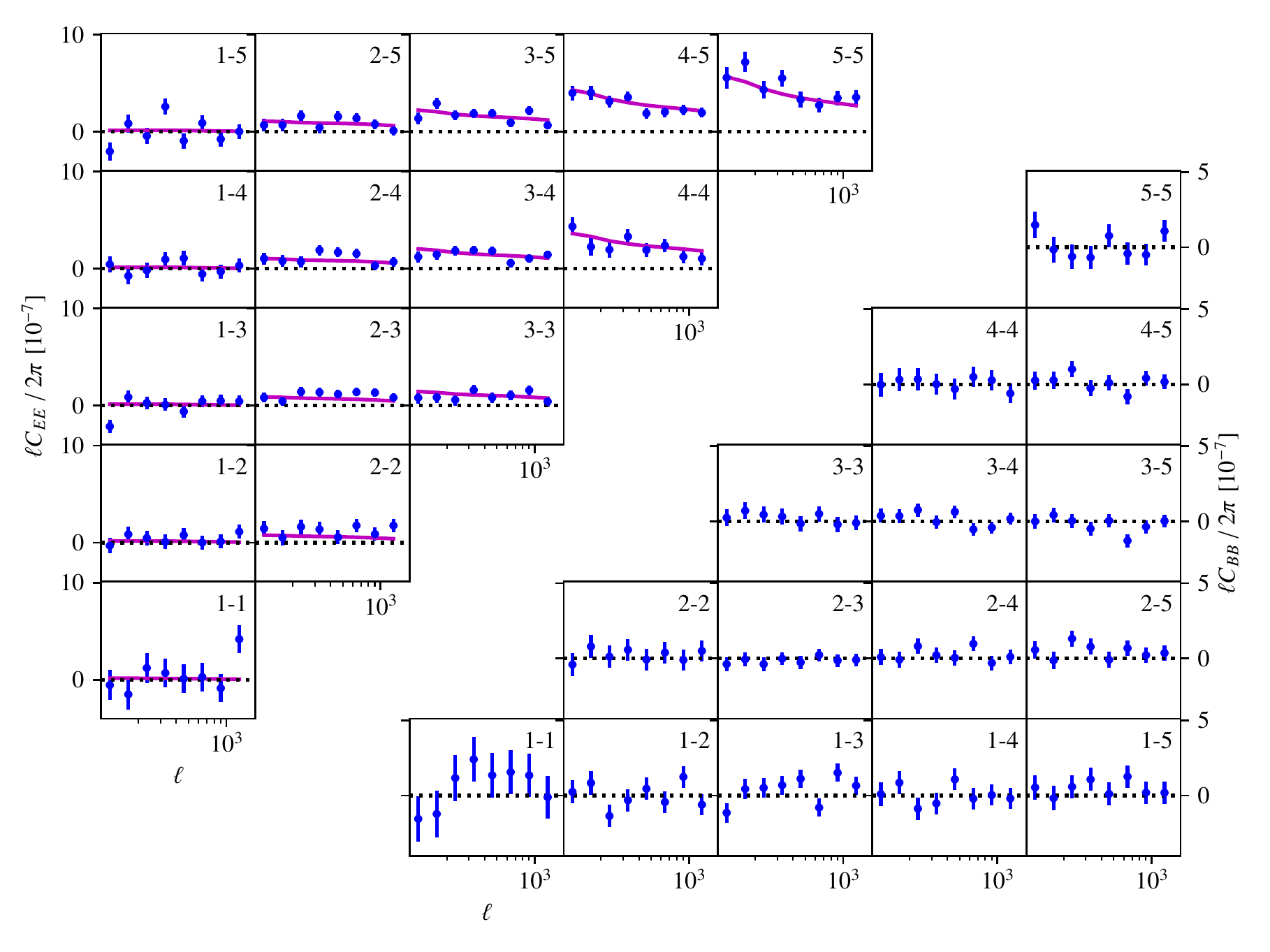}
        \caption{KiDS-1000 cosmic shear power spectra:  Tomographic
          band powers comparing the E-modes (upper left block) with the best-fit
          cosmological model from our combined multi-probe analysis.  The tomographic
        bin combination is indicated in the upper right corner of each
      sub-panel.  The null-test B-modes (lower right block - note the reduced ordinate scale), are
      consistent with zero for both the full data vector and each
     bin combination individually.   The errors are estimated analytically \citep{joachimi/etal:inprep}.  See Sect.~\ref{sec:results} for a discussion on the goodness-of-fit.}
        \label{fig:Pkk}
\end{centering}
\end{figure*}

The observed cosmic shear angular power spectrum, $C_{\epsilon \epsilon}(\ell)$, measures a combination of the distortions arising from weak gravitational lensing by large-scale structures (labelled with a subscript `G') with a low-level contaminating astrophysical signal arising from the intrinsic alignment of galaxies with the large-scale structures within which they are embedded (labelled with a subscript `I').   These contributions can be separated as
\be
\label{eq:cl_cosmicshear}
C^{(ij)}_{\epsilon \epsilon}(\ell) = C^{(ij)}_{\rm GG}(\ell) +
C^{(ij)}_{\rm GI}(\ell) + C^{(ij)}_{\rm IG}(\ell) + C^{(ij)}_{\rm II}(\ell)\;,
\ee
where the indices $i$ and $j$ indicate cross-correlations between the five tomographic source samples.   The theoretical power spectra are given by Limber-approximated projections with
\be
\label{eq:generallimber}
C^{(ij)}_{\rm ab}(\ell) = \int^{\chi_{\rm hor}}_0 \!\!\! \dd \chi\;
\frac{W^{(i)}_{\rm a} (\chi)\; W^{(j)}_{\rm b} (\chi)}{f^2_{\rm
    K}(\chi)}\; P_{\rm m, nl} \br{\frac{\ell+1/2}{f_{\rm K}(\chi)},z(\chi)}\;,
\ee
where ${\rm a,b} \in \bc{\rm I,G}$, $f_{\rm K}(\chi)$ is the comoving angular diameter distance and $\chi$ is the comoving radial distance which runs out to the horizon, $\chi_{\rm hor}$.  The weight functions, $W(\chi)$, encode information about how the signal scales with the KiDS-1000 survey depth \citep[see equations 15 and 16 of][]{joachimi/etal:inprep}.   In the cases of power spectra that include intrinsic `I' terms, the weight function also encodes the intrinsic galaxy alignment model, which we take to be the `NLA' model from \citet{bridle/king:2007}.  For Stage III surveys like KiDS-1000, this model has been shown to be sufficiently flexible, capturing the likely more complex underlying intrinsic alignment model, without biasing cosmological parameters \citep[][see the discussion in Appendix~\ref{app:IAmodel}]{fortuna/etal:2020}.   The cosmological information for cosmic shear power spectrum is contained in both the geometric weight functions, $W(\chi)$, and in the evolution and shape of the non-linear matter power spectrum, $P_{\rm m, nl}(k,z)$, which we model using the halo formalism\footnote{We calculate the non-linear power spectrum using {\sc HMCode} \citep{mead/etal:2016}, which is incorporated in {\sc CAMB} \citep{lewis/bridle:2002}.   \citet{joachimi/etal:inprep} demonstrate that the \citet{mead/etal:2016} halo model prescription provides a sufficiently accurate model of the non-linear matter power spectrum into the highly non-linear regime through a comparison to weak lensing observables emulated using the N-body CosmicEmu simulations \citep{heitmann/etal:2014}.   It also has the added benefit of allowing us to marginalise over our uncertainty on the impact of baryon feedback on the shape of the non-linear total matter power spectrum \citep{semboloni/etal:2011,mead/etal:2015,mead/etal:2020}.} of \citet{mead/etal:2015,mead/etal:2016}.   Weak lensing is therefore a very valuable cosmological probe, as it is sensitive to changes in both the distance-redshift relation and to the growth of structures.

We estimate the cosmic shear angular power spectrum through a linear transformation of the real-space two-point shear correlation function \citep{schneider/etal:2002}.  This approach circumvents the challenge of accurately determining the survey mask for a direct power spectrum estimate.  \citet{joachimi/etal:inprep} detail the apodisation advances that we have adopted for the transformation, in addition to the modelling that we use to account for the minor differences between the theoretical expectation of the true angular power spectrum in Eq.~(\ref{eq:cl_cosmicshear}) and the measured `band powers'.    
 
Fig.~\ref{fig:Pkk} presents the \citet{asgari/etal:inprep} KiDS-1000 cosmic shear power spectra for the auto- and cross-correlated tomographic bins.   Here we have constructed both E-mode (upper left) and B-mode (lower right) band powers in order to isolate any non-lensing B-mode distortions \citep[see equations 17 to 21 of][]{joachimi/etal:inprep}.     As expected from the analysis of \citet{giblin/etal:inprep}, the measured B-modes are found to be consistent with zero\footnote{\citet{giblin/etal:inprep} present a `COSEBIs' B-mode analysis following \citet{asgari/etal:2019}.  The alternative band power B-mode measurement, presented in Fig.~\ref{fig:Pkk}, is consistent with random noise, finding a $p$-value of $p=0.68$ for the full data vector.  Here $p$ corresponds to the probability of randomly producing a noisy B-mode that is more significant than the measurements.  Inspecting each individual tomographic bin combination we find that these are also consistent with random noise with a minimum $p=0.02$ found for the 1-3 bin combination.   A $\sim\!2\,\sigma$ deviation is expected, given the 15 different bin combinations analysed, and we note that the bin combination outlier in this test differs from the $\sim\!2\,\sigma$ deviation bin combination outliers in the two different COSEBIs analyses, supporting the hypothesis that the measured B-modes are simple noise fluctuations.}.   The measured E-modes can be compared to the theoretical expectation from Eq.~(\ref{eq:cl_cosmicshear}), given the best-fit set of cosmological parameters from our multi-probe analysis in Sect.~\ref{sec:results}.

\begin{figure*}
\begin{centering}
        \includegraphics[width=\textwidth]{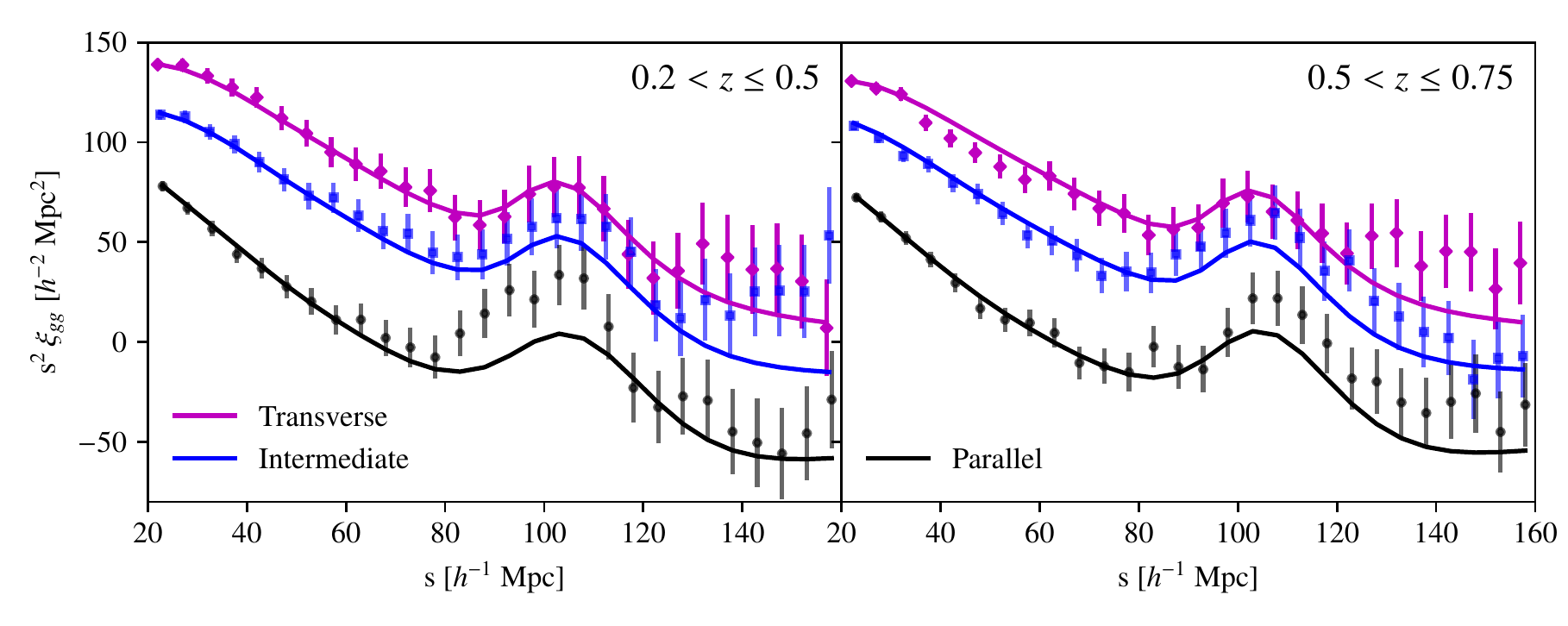}
        \caption{BOSS-DR12 anisotropic clustering from \citet{sanchez/etal:2017}:
          The transverse (pink), intermediate (blue) and parallel
          (black) clustering wedges in two redshift bins, compared 
          with the best-fit
          cosmological model from our combined multi-probe analysis.  The errors, estimated from mock BOSS catalogues \citep{kitaura/etal:2016}, are highly correlated, particularly at large scales (see Fig.~\ref{fig:ttttpcov}).}
        \label{fig:wedges}
        \end{centering}
\end{figure*}

\subsection{Anisotropic galaxy clustering}
\label{sec:clustering}
Galaxy clustering observations probe the 3D non-linear galaxy-galaxy power spectrum and we follow \citet{sanchez/etal:2017} in modelling this quantity based on a perturbation theory approach with
\be
\label{eq:pgg}
P_{\rm gg}(k,z) = \sum_{\alpha,\beta} \alpha\, \beta\, P_{\alpha
  \beta}(k,z) + b_1 \gamma_3^-\, P_{b_1 \gamma_3^-}(k,z) + P_{\rm noise}(k,z)\;.
\ee
Here $\alpha,\beta \in \bb{b_1, b_2, \gamma_2}$, introduce the linear and quadratic bias parameters $b_1$ and $b_2$, in addition to the non-local bias parameters $ \gamma_2$ and $\gamma_3^-$.  Each power spectrum term on the right hand side of the equation is given by different convolutions of the linear matter power spectrum in appendix A of \citet{sanchez/etal:2017}.   In the case of an effective linear galaxy bias model \citep[see for example][]{vanuitert/etal:2018, abbott/etal:2018} only the $b_1$ bias parameter is considered to be non-zero and Eq.~(\ref{eq:pgg}) reduces to $P_{\rm gg}(k,z) =b_1^2 P_{\rm m, nl}^{\rm pert}(k,z)$, where $P_{\rm m, nl}^{\rm pert}(k,z)$ is the perturbation theory estimate of the non-linear matter power spectrum which is accurate at the two percent level to $k \lesssim 0.3 h \,{\rm Mpc}^{-1}$ \citep{sanchez/etal:2017}. 

\citet{sanchez/etal:2017} present the anisotropic redshift-space correlation function of galaxy clustering with the galaxy pairs separated into three `wedges' equidistant\footnote{For the three wedges $i$, the separation in $\mu$ is given by $(i-1)/3 < \mu \leq i/3$.} in $\mu$, where $\mu$ is the cosine of the angle between the line of sight and the line connecting the galaxy pairs.   As such the 3D correlation function is measured for pairs that are either mainly transverse to the line of sight, mainly parallel to the line of sight, or placed into an intermediate sample between these two cases.  The redshift-space correlation function $\xi_{\rm gg} \br{s,\mu,z}$, where $s$ is the co-moving galaxy-pair separation, is given by
\eqa{
\label{eq:bosswedgefouriertrafo}
 \xi_{\rm gg} \br{s,\mu,z} &= \sum_{l=0}^{2} {\rm L}_{2l}(\mu) \frac{(-1)^l
   (4l+1)}{(2 \pi)^2} \int_0^\infty \dd k\, k^2 {\rm j}_{2l}(ks) \\ \nonumber
& \times \int_{-1}^1
 \dd \mu_1 {\rm L}_{2l}(\mu_1) P_{{\rm gg}, s}(k,\mu_1,z)\;,
}
where ${\rm L}_i$ denotes the Legendre polynomial of degree $i$, ${\rm j}_i$ is the spherical Bessel function of order $i$, and $P_{{\rm gg}, s}(k,\mu,z)$ is the 3D redshift-space power spectrum that includes the non-linear real-space power spectrum, Eq.~(\ref{eq:pgg}), and the galaxy-velocity and velocity-velocity power spectrum \citep[see][for details, including how the Alcock-Paczynski distortions are accounted for in the modelling]{sanchez/etal:2017}.     The same model and `wedge' approach was adopted in the Fourier-space anisotropic galaxy clustering analysis of \citet{grieb/etal:2017}, finding consistent results.

Fig.~\ref{fig:wedges} presents the \citet{sanchez/etal:2017} BOSS-DR12 anisotropic clustering correlation functions in three wedges and two redshift slices\footnote{We do not include the \citet{sanchez/etal:2017} central redshift bin measurements in this analysis.  The central bin fully overlaps with the two primary redshift bins, shown in Fig.~\ref{fig:wedges}, and was found not to add any significant constraining power.}, with $0.2<z\leq0.5$, and $0.5<z\leq0.75$, for the scales used in this analysis with $20  < s < 160\, h^{-1}\, {\rm Mpc}$.   The measured correlation functions can be compared to the theoretical expectation\footnote{The `wedge' $\mu$-averaging is given explicitly in equation 1 of \citet{sanchez/etal:2017}.   The perturbative computations for the galaxy-galaxy power spectrum in Eq.~\ref{eq:pgg} are evaluated at a single, effective redshift that is then appropriately scaled to the redshifts of the two redshift bins, following \citet{sanchez/etal:2017}.  Even with this approximation this term is the primary bottleneck of each likelihood evaluation, with a runtime in excess of that of {\sc CAMB}.}  given the best-fit set of cosmological parameters from our joint multi-probe analysis in Sect.~\ref{sec:results}.

\subsection{Galaxy-galaxy lensing}
\label{sec:GGL}

\begin{figure*}
\begin{centering}
        \includegraphics[width=0.95\textwidth]{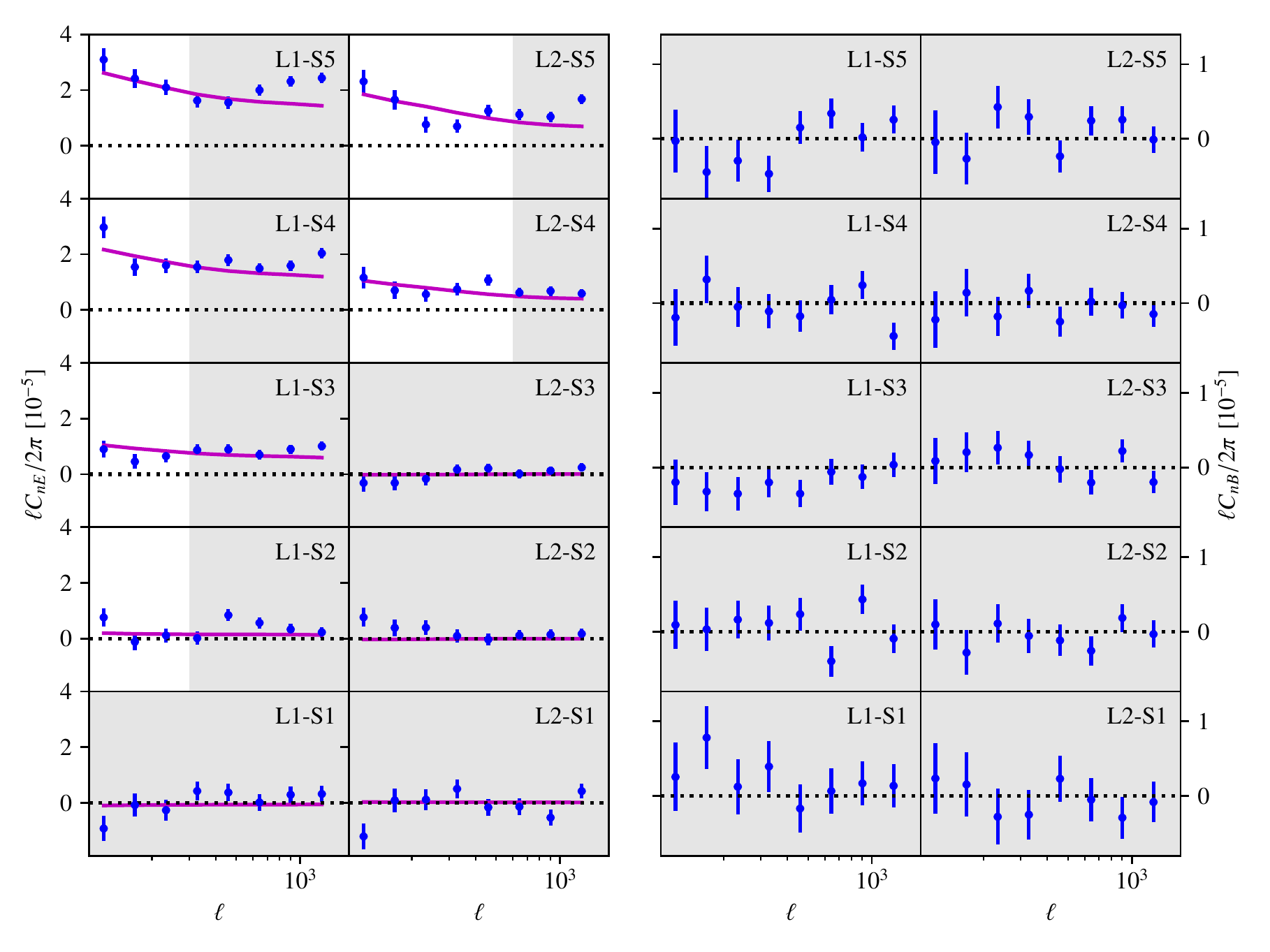}
        \caption{KiDS-1000 galaxy-galaxy lensing power spectra:
          Tomographic band powers comparing the E-modes (left block)
          with the best-fit
          cosmological model from our combined multi-probe analysis.  The tomographic 
        bin combination of BOSS and 2dFLenS lenses (L) with KiDS-1000
        sources (S), is indicated in the upper right corner of each
        sub-panel.  Data within grey regions are not included in the cosmological analysis.
        The null-test B-modes (right block - note the reduced ordinate scale), are
      consistent with zero for both the full data vector and each
     bin combination individually.  The errors are estimated analytically \citep{joachimi/etal:inprep}.}
        \label{fig:Pgk}
        \end{centering}
\end{figure*}

The observed galaxy-galaxy lensing angular power spectra, $C_{\rm n \epsilon}(\ell)$, measure a combination of weak lensing distortions around foreground galaxies (labelled with a subscript `gG') with a low-level intrinsic alignment signal arising from the fraction of the source galaxy population that reside physically close to the lenses (labelled with a subscript `gI').   We also consider the low-level lensing-induced magnification bias (labelled with a subscript `mG').   These three contributions can be separated as 
\be
\label{eq:cl_ggl}
C^{(ij)}_{\rm n \epsilon}(\ell) = C^{(ij)}_{\rm gG}(\ell) +
C^{(ij)}_{\rm gI}(\ell) + C^{(ij)}_{\rm mG}(\ell)  \;,
\ee
where the index $i$ indicates the two lens galaxy samples and the index $j$ indicates the five tomographic source samples.   The theoretical power spectra are given by Limber-approximated projections following Eq.~(\ref{eq:generallimber}), with two key differences.   The first is that the lens weight function is replaced by the redshift distribution of the lenses.   The second is that the non-linear matter power spectrum is replaced with the non-linear cross power spectrum between the galaxy and matter distribution $P_{\rm gm}(k,z)$ \citep[see equations 28 and 29 of][for the full expressions]{joachimi/etal:inprep}.   For the magnification bias power spectrum, $C_{\rm mG}(\ell)$, we refer the reader to appendix B in \citet{joachimi/etal:inprep}.   We find that the inclusion or exclusion of this term has a negligible impact on our cosmological constraints, but we retain it nevertheless.

We adopt the non-linear galaxy bias model from \citet{sanchez/etal:2017}, and approximate the non-linear cross power spectrum as
\eqa{
\label{eq:pgm}
P_{\rm gm}(k,z) &= b_1 P_{\rm m, nl}(k,z) + \left\{ b_2\, {\cal F}_{b_2}(k)
  - \gamma_2\, {\cal F}_{\gamma_2}(k) \right.\\ \nonumber
& \left. -\, \gamma_3^-\, {\cal F}_{\gamma_3^-}(k)
 \right\} P_{\rm m, lin}^{2}(k,z)\;.
}
Here $P_{\rm m, nl}(k,z)$ is the non-linear matter power spectrum modelled using \citet{mead/etal:2016}, in contrast to the less accurate perturbation theory estimate $P_{\rm m, nl}^{\rm pert}(k,z)$ used in Eq.~(\ref{eq:pgg}). 
The logarithm of the functions ${\cal F}_\alpha(k)$ are second-order polynomial fits that we use to model $P_\alpha(k,z_{\rm ref})/P^2_{\rm m, lin}(k,z_{\rm ref})$, the ratio between the different bias terms in the full perturbation theory model in Eq.~(\ref{eq:pgg}), and the square of the linear matter power spectrum.  This approach permits a reasonable extrapolation of the \citet{sanchez/etal:2017} perturbation model into the non-linear regime beyond $k = 0.3 h \,{\rm Mpc}^{-1}$.   This is necessary in order to carry out the redshift-weighted projection of the 3D model to estimate the 2D galaxy-galaxy lensing observable, Eq. (\ref{eq:generallimber}).  No matter which $\ell$-scales we restrict our analysis to, high-$k$ scales will contribute to all angular scales at some level \citep{joachimi/etal:inprep, asgari/etal:2020}.   This approach also decreases the compute time of the galaxy-galaxy lensing likelihood evaluations, by several orders of magnitude, in comparison to the direct perturbative calculation.  

Fig.~\ref{fig:Pgk} presents the KiDS-1000 galaxy-galaxy lensing power spectra, around lenses from the BOSS and 2dFLenS surveys \citep[see][for the real-space KiDS-1000 galaxy-galaxy lensing measurements for BOSS and 2dFLenS separately]{blake/etal:2020}.   Each panel presents the cross-correlation between each of the five different tomographic source bins, denoted `S', with the two different lens bins, denoted `L' (see Table~\ref{tab:datatab} for details).  Here we have constructed both E-mode (left) and B-mode (right) band powers in order to isolate any non-lensing B-mode distortions.     As expected from the analysis of \citet{giblin/etal:inprep}, the measured B-modes are found to be consistent with zero.   The measured E-modes can be compared to the theoretical expectation given the best-fit set of cosmological parameters from our joint multi-probe analysis in Sect.~\ref{sec:results} in the non-shaded regions.    

The shaded regions in Fig.~\ref{fig:Pgk} are excluded from our analysis for two reasons.   For overlapping lens-source bins (L1 with S1 and L2 with S1 to S3), the intrinsic alignment terms $C_{\rm gI}(\ell)$ are expected to become significant.   This raises the question of the validity of the arguably rudimentary `NLA' intrinsic alignment model when used in combination with a non-linear galaxy bias model \citep[see][for a self-consistent pertubative approach to both intrinsic alignment and galaxy bias modelling]{blazek/etal:2019}.   As these bin combinations carry little cosmological information, we exclude this data from our cosmological inference analysis, using it instead in a redshift-scaling null test of the catalogue in \citet{giblin/etal:inprep}.   For separated lens-source bins we introduce a maximum $\ell$-scale beyond which the contributions from scales $k > 0.3 h \,{\rm Mpc}^{-1}$ become significant\footnote{Figure 2 in \citet{joachimi/etal:inprep}, demonstrates that the $k > 0.3 h \,{\rm Mpc}^{-1}$ scales only contribute to the very low-level oscillating wings of the Fourier-space filters for the $\ell$-scales selected in Fig.~\ref{fig:Pgk}}.  In this regime uncertainties in the extrapolation of the \citet{sanchez/etal:2017} non-linear galaxy bias model into the non-linear regime (Eq.~\ref{eq:pgm}) may well render the $C_{\rm n \epsilon}(\ell)$ model invalid.   The $\ell$-limit depends on the redshift of the lens bin.    Fig.~\ref{fig:Pgk} therefore serves as an important illustration of the necessity of improving non-linear galaxy bias and non-linear intrinsic alignment modelling for future studies, in order to fully exploit the cosmological signal contained within the galaxy-galaxy lensing observable.

\subsection{Multi-probe covariance}
\label{sec:Cov}
\citet{joachimi/etal:inprep} present the multi-probe covariance matrix adopted in this study, verified through an analysis of over $20\,000$ fast full-sky mock galaxy catalogues derived from lognormal random fields.   Given that only 4\% of the BOSS footprint overlaps with KiDS-1000, in an initial step we validate the approximation that the BOSS anisotropic galaxy clustering observations are uncorrelated with the cosmic shear and galaxy-galaxy lensing observations.   By imposing realistic overlapping BOSS and KiDS-1000 footprints in our mock catalogues, we find that cross-correlation, between the projected BOSS-like angular galaxy correlation function, and the KiDS-1000-like weak lensing signals, is less than $\sim\!5\%$ of the auto-correlation terms along the diagonal of covariance matrix.  With such a low cross-correlation, we can safely assume independence between the clustering and lensing observations, allowing us to adopt the \citet{sanchez/etal:2017} covariance matrix\footnote{The BOSS $\xi_{\rm gg}$ covariance is derived from the {\sc MD-Patchy} BOSS mock catalogues of \citet{kitaura/etal:2016}.} for the anisotropic galaxy clustering observations, $\xi_{\rm gg}(s,\mu,z)$, setting the clustering-lensing cross-correlation terms to zero.   

The covariance of the two weak lensing observations is calculated analytically, combining terms that model pure Gaussian shape noise, survey sampling variance, and the noise-mixing that occurs between these two components, in addition to higher-order terms that account for mode-mixing between the in-survey modes and between the observed in-survey and the unobserved out-of-survey modes \citep[known as super-sample covariance,][]{takada/hu:2013}. 
The covariance also includes a contribution to account for our uncertainty on the multiplicative shear calibration correction \citep{kannawadi/etal:2019}. \citet{joachimi/etal:inprep} demonstrate that every term in the covariance is important, each dominating in different regions of the covariance with one exception: non-Gaussian variance between the in-survey modes is always sub-dominant.  We therefore review the approximations made in these analytical calculations.   
Whilst the complex KiDS-1000 mask is fully accounted for in the shape-noise terms, and the super-sample terms, for all other terms it is assumed that the scales we measure are much smaller than any large-scale features in the survey footprint. 
Furthermore, the survey is assumed to be homogeneous in its depth, which is invalid for any ground-based survey where the survey depth becomes a sensitive function of the observing conditions \citep{heydenreich/etal:2020}.   With mock catalogues, we have the freedom to impose complex masks and variable depth to quantify the impact of these effects on the derived covariance, finding differences typically $\lesssim 10\%$, with a maximum difference of $\sim\!20 \%$.  The majority of the differences were found to be driven by the mix-term between the Gaussian shape noise and the Gaussian sampling variance.    Through a mock multi-probe data vector inference analysis, \citet{joachimi/etal:inprep} demonstrate that these differences in the covariance are not expected to lead to any systematic bias in the recovery of the KiDS-1000 cosmological parameters, nor to any significant differences in the confidence regions of the recovered parameters.   We therefore adopt an analytical covariance in our analysis,  shown in Fig.~\ref{fig:ttttpcov}, and refer the reader to \citet{joachimi/etal:inprep} for further details, where their section 4 presents the mocks, section 5 and appendix E presents the analytical covariance model, and appendix D presents detailed comparisons of the mock and analytical covariance.

\subsection{Parameter inference methodology}
\label{sec:KCAP}
We use the KiDS Cosmology Analysis Pipeline, {\sc KCAP}\footnote{{\sc KCAP}: \url{https://github.com/KiDS-WL/KCAP}} built from the {\sc CosmoSIS} analysis framework of \citet{zuntz/etal:2015}, adopting the nested sampling algorithm {\sc Multinest} \citep{feroz/hobson:2008,feroz/etal:2009,feroz/etal:2019}.  The {\sc KCAP} bespoke modules include: the BOSS wedges likelihood from \citet{sanchez/etal:2017}; the band power cosmic shear and galaxy-galaxy lensing likelihood based on Eq.~(\ref{eq:cl_cosmicshear}) and Eq.~(\ref{eq:pgm});  tools to permit correlated priors on nuisance parameters; and tools to sample over the clustering parameter $S_8= \sigma_8 \sqrt{\Omega_{\rm m}/0.3}$, a parameter which is typically only derived.  Scripts are also provided to derive the best-fit parameter values at the maximum multivariate posterior, denoted MAP (maximum a posteriori), and an associated credible region given by the projected joint highest posterior density region, which we denote PJ-HPD \citep{joachimi/etal:inprep}.  This concise list of new modules reflect the primary updates in the KiDS-1000 parameter inference methodology compared to previous KiDS analyses, which we discuss in more detail below.

Our \tttp model has 20 free parameters, with five to describe flat $\Lambda$CDM in addition to fifteen nuisance parameters.   Eight of these nuisance parameters describe the galaxy bias model, with four in each lens redshift bin. The remaining seven allow us to marginalise over our uncertainty on the impact of baryon feedback (one parameter), intrinsic galaxy alignment (one parameter), and the mean of the source redshift distribution in each tomographic bin (five correlated parameters).   The priors adopted for each parameter are listed in Appendix~\ref{app:priors}.  

Adopted priors are usually survey-specific, with the intention to be uninformative on the parameter that lensing studies are most sensitive to, $S_8$.  
Different prior choices, particularly on the amplitude of the primordial power spectrum of scalar density fluctuations, $A_{\rm s}$, have however been shown to lead to non-negligible changes in the derived $S_8$ parameter \citep{joudaki/etal:2017,chang/etal:2019, joudaki/etal:2020, asgari/etal:2020_KD}. 
\citet{joachimi/etal:inprep} show that even with wide priors on $A_{\rm s}$,  the sampling region in the $\sigma_8$-$\Omega_{\rm m}$ plane is significantly truncated at low values of $\sigma_8$ and $\Omega_{\rm m}$, with the potential to introduce a subtle bias towards low values of $\sigma_8$.   In this analysis, we address this important issue of implicit informative priors by sampling directly in $S_8$.   By adopting a very wide $S_8$ prior, our constraints on $S_8$ are therefore not impacted by our choice of prior.  We note, however, that this approach is expected to lead to a more conservative constraint on $S_8$, compared to an analysis that adopts a uniform prior\footnote{For quantitative information about the impact of implicit $A_{\rm s}$ priors, see table 5 and figure 22 of \citet{joachimi/etal:inprep}.} on $\ln A_{\rm s}$.

We account for the uncertainty in our source redshift distributions using nuisance parameters, $\delta_z^i$, which modify the mean redshift of each tomographic bin $i$.   By analysing mock KiDS catalogues, \citet{wright/etal:2020} determined the mean bias per redshift bin, $\mu^i$, and also the covariance between the different redshift bins, $C_{\delta z}$.   This covariance arises from sampling variance in the spectroscopic training sample, which impacts, to some degree, the redshift calibration of all bins.    We therefore adopt the multivariate Gaussian prior ${\cal N}(\vek{\mu};\vek{C}_{\delta z})$, for the vector $\vek{\delta}_z$ \citep[see section 3 of][for details]{hildebrandt/etal:inprep}. 

Adopting the Bayesian paradigm for inference, we provide our constraints in the form of a series of samples that describe the full posterior 
distribution\footnote{Our {\sc Multinest} full posterior samples can be accessed at \href{http://kids.strw.leidenuniv.nl/DR4/KiDS-1000_3x2pt_Cosmology.php}{kids.strw.leidenuniv.nl/DR4/KiDS-1000\_3x2pt\_Cosmology.php}}.  
In Sect.~\ref{sec:results} we explore this multi-dimensional posterior in the traditional way, visualising the 2D and 1D marginal posterior distributions for a selection of parameters.  
In cosmological parameter inference it is standard to also report a point estimate of the one-dimensional marginal posterior distribution with an associated 68\% credible interval.   It
is not always stated, however, how these point estimates and intervals are defined.   

We provide two different point estimates for our cosmological parameters, with the first reporting the standard maximum of the marginal distribution, 
along with a credible interval that encompasses 68\% of the marginal highest posterior density, which we denote by M-HPD.  
For the high-dimensional parameter space of a multi-probe weak lensing analysis, we find that this standard marginalised point estimate leads to a value for $S_8$ that is lower than the 
maximum of the multivariate joint posterior, with an offset of up to $\sim\!1\,\sigma$, dependent on which probes are combined \citep[see section 7 of][]{joachimi/etal:inprep}.
This is not a result of an error in the {\sc KCAP} inference pipeline. 
Rather, it is a generic feature of projecting high-dimensional asymmetric distributions into one dimension, prompting the development of an alternative approach to reporting point estimates for cosmological parameters.

Our fiducial $S_8$ constraints follow this alternative, reporting the parameter value at the maximum of the joint posterior (MAP), along with a 68\% credible interval based on the joint, multi-dimensional highest posterior density region, projected onto the marginal posterior of the $S_8$ parameter (PJ-HPD).   
Here we step through the posterior {\sc Multinest} samples, ordered by their decreasing posterior density.  
For each model parameter we determine the extrema within the $n$ highest posterior samples, and the posterior mass contained within the marginal distribution of each parameter, limited by the extrema values.   
We iterate, increasing the number of samples analysed, $n$, until the posterior mass reaches the desired 68\% level.   
The PJ-HPD credible interval is then reported as the parameter extrema at this point $n$ in the sample list, and we repeat the process for each model parameter of interest \citep[see section 6.4 of][for further details]{joachimi/etal:inprep}.   

We note that the MAP reported by {\sc Multinest} provides a noisy estimate of the true MAP due to the finite number of samples, and we therefore conduct an optimisation step using the 36 samples with the highest posterior values as starting points.
We use both the \citet{nelder/mead:1965} and \citet{Powell1964} optimisation algorithms, as well as two-step optimisation using both algorithms.  
While the MAP estimates found in this optimisation step increase the posterior probability by a factor of 2 to 4 compared to the MAP estimated from the {\sc Multinest} samples, they exhibit scatter in parameters constrained by the galaxy clustering likelihood. 
We suspect this is due to numerical noise in the galaxy clustering likelihood which results in many local minima that inhibit the convergence of the optimisation step. 
This suspicion is strengthened by the fact that the MAP estimates do not exhibit this scatter for probe combinations that exclude the clustering observable. 
For this reason, we report the median of the MAP estimates, weighted by their posterior probability, for probes that include the galaxy clustering likelihood, since a global optimisation would be computationally prohibitively expensive. 
For the other probes the reported MAP is given by the parameter set at the maximum posterior found amongst all estimates \citep[see also][who adopt a similar approach]{muir/etal:2020}. 

We note that the presence of offsets between marginal $S_8$ constraints and those derived from the full multivariate joint posterior 
highlights how efforts to accurately quantify tension based solely on one-point estimates should be undertaken with some level of caution.  
Tension can also be assessed in terms of the overlap between the full posterior distributions \citep[see for example][]{handley/lemos:2019,lemos/etal:2019,Raveri2019}, which we discuss further in Sect.~\ref{sec:planck_comp} and Appendix~\ref{app:tensionest}.

\section{Results}
\label{sec:results}
\begin{figure}
	\begin{center}
		\includegraphics[width=\columnwidth]{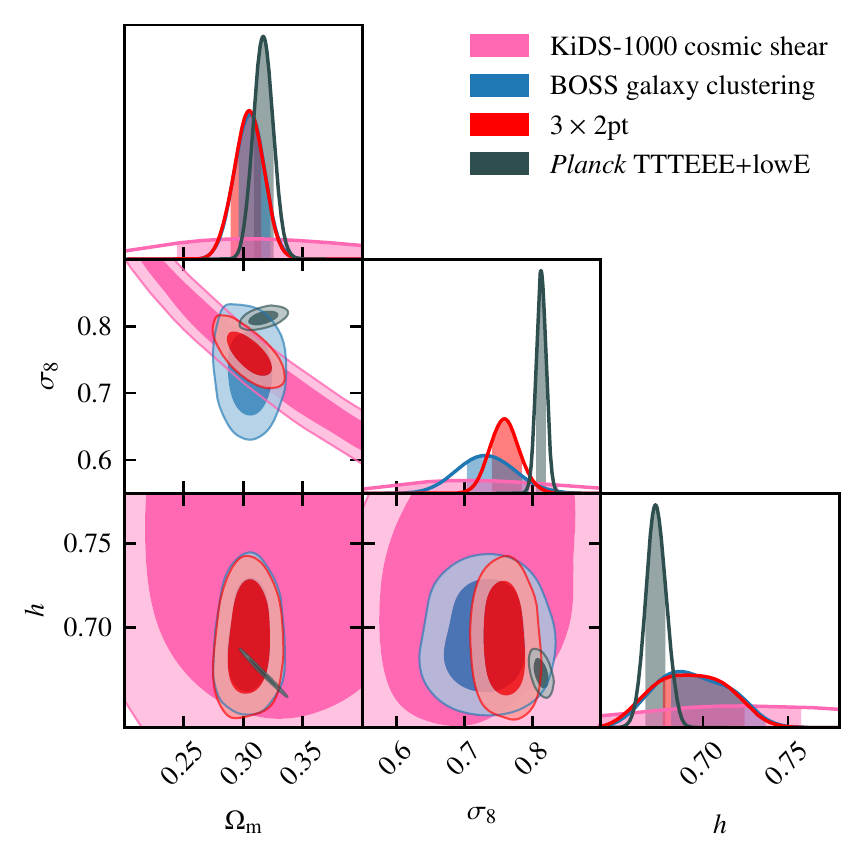}
		\caption{Marginal multi-probe constraints on the flat $\Lambda$CDM cosmological model, for the matter fluctuation amplitude parameter, $\sigma_8$, the matter density parameter, $\Omega_{\rm m}$, and the dimensionless Hubble parameter, $h$.  The BOSS galaxy clustering constraints (blue), can be compared to the KiDS-1000 cosmic shear constraints (pink), the combined $3\times2{\rm pt}$ analysis (red), and CMB constraints from \citet[][grey]{planck/etal:2018}.}
		\label{fig:cosmology-params}
	\end{center}
\end{figure}

We present our multi-probe constraints on the cosmological parameters of the flat $\Lambda$CDM model in Fig.~\ref{fig:cosmology-params}, showing the marginalised posterior distributions for matter fluctuation amplitude parameter, $\sigma_8$, the matter density parameter, $\Omega_{\rm m}$, and the dimensionless Hubble parameter, $h$, where the BOSS galaxy clustering constraints (shown blue), break the $\sigma_8$-$\Omega_{\rm m}$ degeneracy in the KiDS-1000 cosmic shear constraints (shown pink), resulting in tight constraints on $\sigma_8$ in the combined \tttp analysis (shown red). 
Reporting the MAP values with PJ-HPD credible intervals for the parameters that we are most sensitive to, we find 
\eqa{
\sigma_8 &= \ksigmaeightval \\ \nonumber
\Omega_{\rm m} &= \kOmegamval \\ \nonumber
S_8 &= \kSeightval \, .
}
Our constraints can be compared to the marginalised posterior distributions from {\it Planck} (shown grey in Fig.~\ref{fig:cosmology-params}), finding consistency between the marginalised constraints on $\Omega_{\rm m}$ and $h$, but an offset in $\sigma_8$,  which we discuss in detail in Sect.~\ref{sec:planck_comp}.

\begin{figure}
	\begin{center}
		\includegraphics[width=\columnwidth]{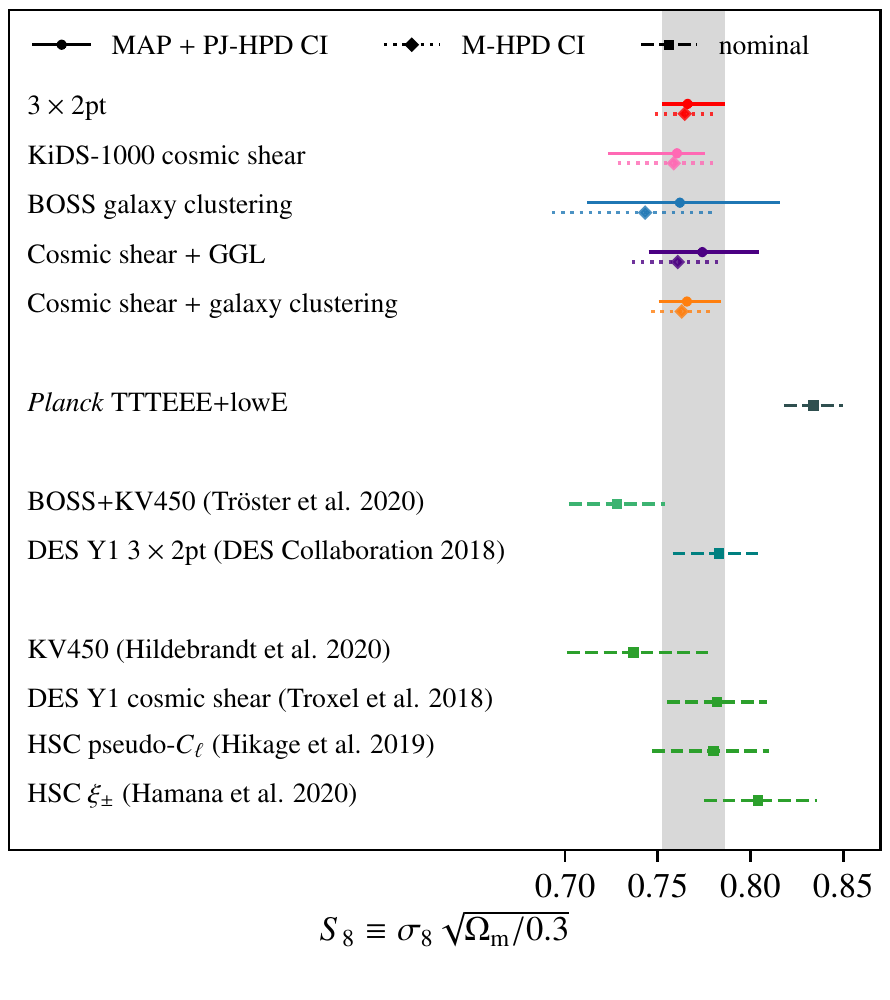}
		\caption{Constraints on the structure growth parameter $S_{8} = \sigma_8 \sqrt{\Omega_{\rm m}/0.3}$ for different probe combinations: $3\times2$pt, KiDS-1000 cosmic shear, BOSS galaxy clustering, cosmic shear with galaxy-galaxy lensing (GGL), and cosmic shear with galaxy clustering.   Our fiducial and preferred MAP with PJ-HPD credible interval (solid) can be compared to the standard, but shifted, marginal posterior mode with M-HPD credible intervals (dotted).    Our results can also be compared to weak lensing measurements from the literature, which typically quote the mean of the marginal posterior mode with tail credible intervals (dashed). 
		\label{fig:S8comp}}
	\end{center}
\end{figure}

Tabulated constraints for the full set of cosmological parameters are presented in Appendix~\ref{app:parameter-constraints}, quoting our fiducial MAP with PJ-HPD credible intervals along with the marginal posterior mode with M-HPD credible intervals. 
As discussed in \citet{joachimi/etal:inprep}, the marginal mode estimate is known to yield systematically low values of $S_8$ in mock data analyses.   This effect can be seen in Fig.~\ref{fig:S8comp} which compares the joint posterior constraints (solid) with the marginal posterior constraints (dashed).  

\begin{figure*}
	\begin{center}
		\includegraphics[width=\textwidth]{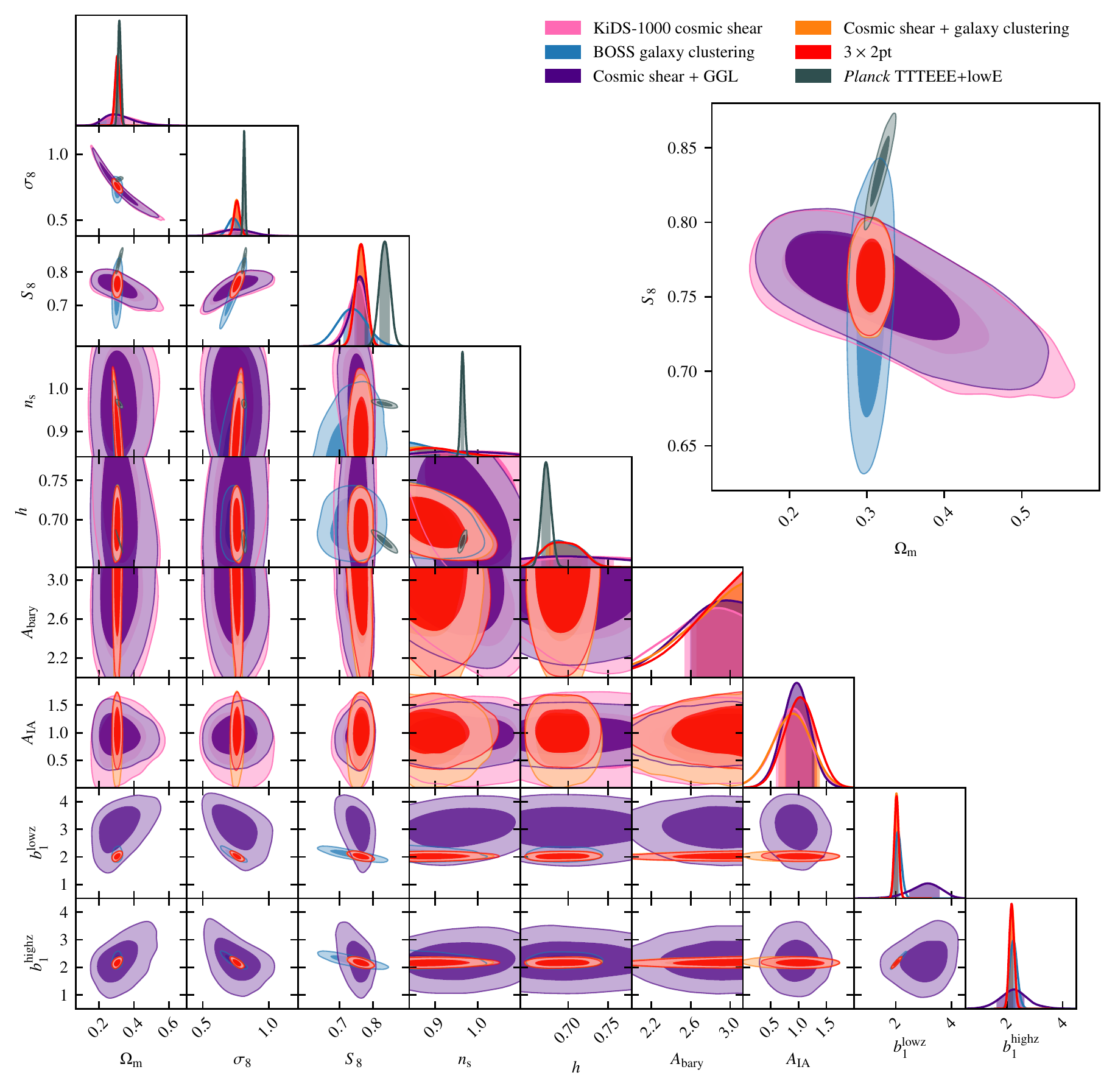}
		\caption{Marginalised posterior distributions for an extended set of cosmological parameters covering the matter density parameter, $\Omega_{\rm m}$, the matter fluctuation amplitude parameter, $\sigma_8$, the structure growth parameter, $S_8$, the spectral index, $n_{\rm s}$, the dimensionless Hubble parameter, $h$, the baryon feedback amplitude parameter, $A_{\rm bary}$, the intrinsic alignment amplitude, $A_{\rm IA}$, and the linear bias parameters for the low and high BOSS redshift bins, $b_1$.   The KiDS-1000 cosmic shear results (pink), can be compared to the BOSS galaxy clustering results (blue), the combination of cosmic shear with BOSS and 2dFLenS galaxy-galaxy lensing (GGL, purple), and the full $3\times2$pt analysis (red).  The combination of cosmic shear with galaxy clustering (orange) is only distinguishable from the \tttp result in the $A_{\rm bary}$ and $A_{\rm IA}$ panels.  For parameters constrained by the CMB, we also include constraints from \citet[][grey]{planck/etal:2018}.}
		\label{fig:cosmology-params-all}
	\end{center}
\end{figure*}

We find good agreement between the different probe combinations and single-probe $S_8$ constraints, demonstrating internal consistency between the different cosmological probes, in Fig.~\ref{fig:S8comp}.  
As forecast by \citet{joachimi/etal:inprep}, the addition of the galaxy-galaxy lensing observable adds very little constraining power, with similar results found for the full \tttp analysis and the combined cosmic shear and clustering analysis. 
This is primarily a result of the significant full area of BOSS in comparison to the size of the BOSS-KiDS overlap region.   The lack of an accurate galaxy bias model on the deeply non-linear scales that weak lensing probes also prohibits the inclusion of large sections of our galaxy-galaxy lensing data vector, shown in Fig.~\ref{fig:Pgk}.  
The addition of the galaxy-galaxy lensing does, however, serve to moderately tighten constraints on the amplitude of the intrinsic alignment model $A_{\rm IA}$,  as seen in Fig.~\ref{fig:cosmology-params-all}. 

Fig.~\ref{fig:S8comp} also demonstrates the good agreement between our constraints and weak lensing results from the literature, comparing to cosmic shear-only results from the Hyper Suprime-Cam Strategic Programme \citep[HSC,][]{hikage/etal:2019,hamana/etal:2020}, DES Y1 \citep{troxel/etal:2018} and an earlier KiDS analysis \citep[KV450][]{hildebrandt/etal:2020}, in addition to the previous KV450-BOSS `$2\times2$pt' analysis of \citet{troester/etal:2020} and the DES Y1 \tttp analysis from \citet{abbott/etal:2018}.   We refer the reader to \citet{asgari/etal:inprep} for a discussion and comparison of different cosmic shear results.  In Sect.~\ref{sec:WL_comp} we present a more detailed comparison of our results with \tttp results in the literature.

Fig.~\ref{fig:cosmology-params-all} displays the marginal posterior distributions for an extended set of cosmological parameters.  
We find that the allowed range for the linear galaxy bias,  $b_1$, in each redshift bin (lower two rows), is almost halved with the addition of the weak lensing data. 
This constraint does not arise, however, from the sensitivity of the galaxy-galaxy lensing observable to galaxy bias (shown to be relatively weak in the purple cosmic shear + GGL contours). 
Instead, in this analysis, it is a result of the degeneracy breaking in the $\sigma_8$-$\Omega_{\rm m}$ plane, tightening constraints on $\sigma_8$ which, for galaxy clustering, is degenerate with galaxy bias. 
The improved constraints on galaxy bias do not, however, fold through to improved constraints on $h$, which the weak lensing data adds very little information to. 

\begin{table*}
	\begin{center}
		\caption{Goodness-of-fit of the flat $\Lambda$CDM cosmological model to each of the single and joint probe combinations with cosmic shear, galaxy clustering and galaxy-galaxy lensing (GGL).}
		\label{tab:goodness-of-fit}
\begin{tabular}{lcccccc}
    \toprule
    Probe             & $\chi^2_{\rm MAP}$  & Data DoF  & Model DoF                   & PTE  & Model DoF          & PTE    \\
                      &                     &           &\citep{joachimi/etal:inprep} &      & \citep{Raveri2019} & \\
    \midrule
	KiDS-1000 cosmic shear     & $152.1$ & $120$  &4.5 & 0.013 &3.0 & 0.016 \\
	BOSS galaxy clustering & $167.7$ & $168$  &-- & -- &10.6 & 0.272 \\
	Cosmic shear + GGL & $178.7$ & $142$  &8.7 & 0.005 &7.3 & 0.007 \\
	Cosmic shear + galaxy clustering & $319.9$ & $288$  &-- & -- &11.9 & 0.036 \\
	$3\times2$pt & $356.2$ & $310$  &-- & -- &12.5 & 0.011 \\

    \bottomrule
\end{tabular}
	\end{center}
	\tablefoot{We list the $\chi^2$ value at the maximum of the posterior, the number of degrees of freedom (DoF) of the data, the effective DoF of the model, and the probability to exceed (PTE) the measured $\chi^{2}$ value, assuming the total DoF are given by $\text{data DoF}-\text{model DoF}$. 
The effective DoF of the model are estimated following \citet{joachimi/etal:inprep} and \citet{Raveri2019}, accounting for the impact of priors and non-linear dependencies between the parameters. }
\end{table*}

For our primary cosmological parameter, $S_8$, our constraints are uninformed by our choice of priors.    This statement cannot be made for the other $\Lambda$CDM parameters, however, as shown in Fig.~\ref{fig:cosmology-params-all}.   The most informative prior that we have introduced to our \tttp analysis is on the spectral index, $n_{\rm s}$.  As noted by \citet{troester/etal:2020}, the BOSS galaxy clustering constraints favour a low value for $n_{\rm s}$, where they find $n_{\rm s} = 0.815 \pm 0.085$. 
From the \citet{troester/etal:2020} sensitivity analysis to the adopted maximum clustering scale we observe that this preference appears to be driven by the amplitude of the large scale clustering signal with $s > 100 \, h^{-1}\, {\rm Mpc}$.  We note that spurious excess power in this regime could plausibly arise from variations in the stellar density impacting the BOSS galaxy selection function \citep{ross/etal:2017}.  Our choice to impose a theoretically motivated informative prior for $n_{\rm s}$, as listed in Table~\ref{tab:priors}, helps to negate this potential systematic effect without degrading the overall goodness-of-fit to the galaxy clustering measurements.  Our prior choice is certainly no more informative than the $n_{\rm s}$ priors that are typically used in weak lensing and clustering analyses \citep[see for example][]{abbott/etal:2018,eBOSS/etal:2020}. 
We recognise, however, that this well-motivated prior choice acts to improve the BOSS-only error on $\Omega_{\rm m}$ by roughly a third, and decrease the BOSS-only best-fitting value for $\Omega_{\rm m}$ and $h$ by $\sim\! 0.5\,\sigma$ (see Fig.~\ref{fig:ns-prior}).  With $<10\%$ differences on the constraints on $S_8$ and $h$, however, and only a $\sim\! 0.1\,\sigma$ difference in the BOSS-only best-fitting value for $S_8$, which is consistent with the typical variation between different {\sc Multinest} analyses, we conclude that our prior choice does not impact on our primary $S_8$ constraints (see Appendix~\ref{app:priors}).   With the informative or uninformative $n_{\rm s}$ prior, our constraints on $h$ remain consistent with the Hubble parameter constraints from both \citet{planck/etal:2018} and \citet{riess/etal:2019}.

\begin{figure}
	\begin{center}
		\includegraphics[width=\columnwidth]{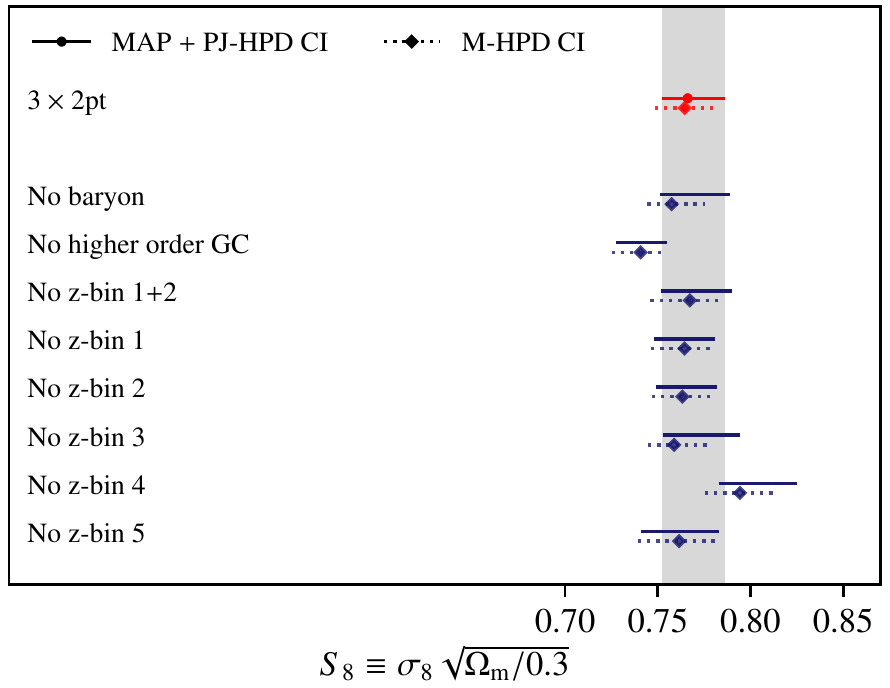}
		\caption{\tttp constraints on $S_8$ for a series of sensitivity tests; when we ignore the impact of baryon feedback (the `No baryon' case), limit the analysis to a linear galaxy bias model (the `No higher order GC' case), and remove individual tomographic bins from our weak lensing observables.  
		\label{fig:S8comp_sensitivity}}
	\end{center}
\end{figure}

Fig.~\ref{fig:S8comp_sensitivity} illustrates the results of a series of sensitivity tests, where we explore how our \tttp constraints on $S_8$ change when: 
we ignore the impact of baryon feedback (the `No baryon' case) by fixing $A_{\rm bary}=3.13$, corresponding to the non-linear matter power spectrum for a dark-matter only cosmology; 
we limit the analysis to a linear galaxy bias model, setting all higher-order bias terms in Eq. (\ref{eq:pgg}) to zero, as well as restricting the redshift-space distortion model to a Gaussian velocity distribution; 
and we remove individual tomographic bins from our weak lensing observables. 
The systematic offset that arises from the use of a linear-bias model highlights the importance of accurate non-linear galaxy bias modelling in \tttp analyses.     
This series of tests is dissected further in Appendix~\ref{app:sensitivity}, and complements the detailed KiDS-1000 
internal consistency analysis of \citet[][appendix B]{asgari/etal:inprep}, 
which demonstrates that the change seen with the removal of tomographic bin 4 is fully consistent with expected statistical fluctuations.

Table~\ref{tab:goodness-of-fit} records the goodness-of-fit for each component in our \tttp analysis, where we report the $\chi^2$ value at the maximum posterior, $\chi^2_{\rm MAP}$ (see Sect.~\ref{sec:KCAP} for a discussion of our optimised MAP-finder).  
The effective number of degrees of freedom (DoF) does not equate to the standard difference between the total number of data points (Data DoF) and the total number of model parameters (20 in the case of our \tttp analysis), as a result of the adopted priors and the non-linear dependencies that exist between the model parameters.   For some probe combinations we calculate the effective number of degrees of freedom in the model (Model DoF), using the estimator described in section 6.3 of \citet{joachimi/etal:inprep}.    
As this approach is computationally expensive, however, we also estimate the Model DoF following \citet{Raveri2019}, recognising that, for the cases explored in \citet{joachimi/etal:inprep}, this approach results in a slightly lower model DoF.

We find that the goodness-of-fit is excellent for the BOSS galaxy clustering.  For all other cases, the goodness-of-fit is certainly acceptable\footnote{We define acceptable as the PTE $p \geq 0.001$, which corresponds to less than a $\sim\!3\,\sigma$ event.   \citet{abbott/etal:2018} define acceptable as $\chi^2/{\rm DoF} < 1.4$.  We meet both these requirements.}, with the probability to exceed the measured $\chi^2$ given by $p \gtrsim 0.01$.
We note that the cosmic shear analysis of \citet{asgari/etal:inprep} shows no significant changes in the inferred cosmological parameters when using different two-point statistics which exhibit an excellent goodness-of-fit.    As such, we could be subject to an unlucky noise fluctuation that particularly impacts the band power estimator in Eq. (\ref{eq:cl_cosmicshear}).  Cautiously inspecting Fig.~\ref{fig:Pkk}, as `$\chi$-by-eye' is particularly dangerous with correlated data points, we nevertheless note a handful of outlying points, for example the low $\ell$-scales in the fifth tomographic bin.   We also note that \citet{giblin/etal:inprep} document a significant but low-level PSF residual systematic in the KiDS-1000 fourth and fifth tomographic bins that was shown to reduce the overall goodness-of-fit in a cosmic shear analysis, but not bias the recovered cosmological parameters \citep[see the discussion in][]{amara/refregier:2008}.  Future work to remove these low-level residual distortions is therefore expected to further improve the goodness-of-fit.

\subsection{Comparison with weak lensing surveys}
\label{sec:WL_comp}
Our results are consistent with weak lensing constraints in the literature.   We limit our discussion in this section to published \tttp analyses, referring the reader to \citet{asgari/etal:inprep} who discuss how the KiDS-1000 cosmic shear results compare with other weak lensing surveys.   We note that direct comparisons of cosmological parameters should be approached with some caution, as the priors adopted by different surveys and analyses are often informative \citep[see section 6.1 in][]{joachimi/etal:inprep}.   Homogenising priors for cosmic shear analyses, for example, has been shown to lead to different conclusions when assessing inter-survey consistency \citep{chang/etal:2019, joudaki/etal:2020, asgari/etal:2020_KD}.   

\begin{figure}
	\begin{center}
		\includegraphics[width=\columnwidth]{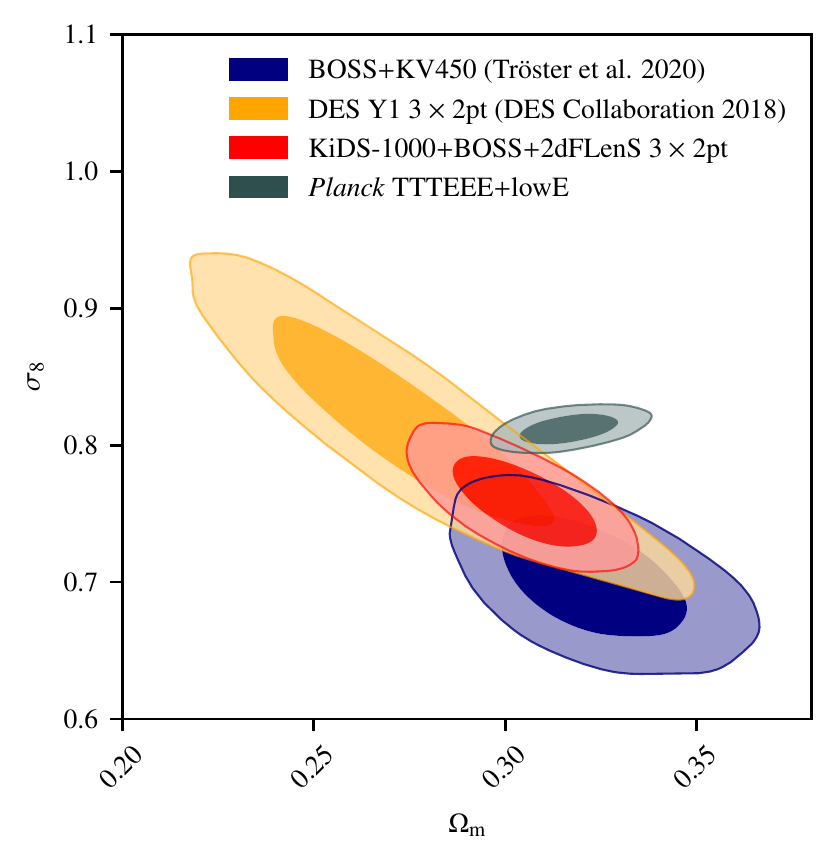}
		\caption{Marginalised posterior distribution in the $\sigma_8$-$\Omega_{\rm m}$ plane, comparing the \tttp analyses from KiDS-1000 with BOSS and 2dFLenS, with the \tttp analysis from DES Y1 \citep{abbott/etal:2018}, and the CMB constraints from \citet{planck/etal:2018}.   The KiDS-1000 \tttp result can also be compared to our previous KV450-BOSS analysis from \citet{troester/etal:2020}. 
		\label{fig:DES_KiDS_comp}}
	\end{center}
\end{figure}

\citet{abbott/etal:2018} present the first year \tttp DES analysis (DES Y1), finding $S_8=0.773^{+0.026}_{-0.020}$, where they report the marginal posterior maximum and the tail credible intervals.  
This is in excellent agreement with our equivalent result, differing by $0.3\,\sigma$, with the DES-Y1 error being 40\% larger than the KiDS-1000-BOSS \tttp results.  The inclusion of BOSS to our \tttp analysis results in tight constraints on $\Omega_{\rm m}$.  
This leads to joint KiDS-1000-BOSS constraints on $\sigma_8=0.760^{+0.021}_{-0.023}$ that are more than twice as constraining compared to the DES Y1-alone \tttp analysis which found $\sigma_8=0.817^{+0.045}_{-0.056}$, as shown in Fig.~\ref{fig:DES_KiDS_comp}. 
This comparison serves to highlight the additional power that can be extracted through the combination of spectroscopic and photometric surveys,  and the promising future for the planned overlap between the Dark Energy Spectroscopic Instrument survey \citep{DESI/etal:2016} and the 4-metre Multi-Object Spectroscopic Telescope \citep[4MOST,][]{richard/etal:2019},
with {\it Euclid} and the Vera C. Rubin Observatory Legacy Survey of Space and Time \citep{laureijs/etal:2011,lsst/etal:2009}, in addition to the nearer-term $\sim\!1400\,\mathrm{deg}^{2}$ of overlap between BOSS and HSC \citep{aihara/etal:2019}. 

\citet{vanuitert/etal:2018} and \citet{joudaki/etal:2018} present \tttp analyses for the second KiDS weak lensing release (KiDS-450), finding, respectively, $S_8 = 0.800_{-0.027}^{+0.029}$ (KiDS with GAMA) and $S_8 = 0.742 \pm 0.035$ (KiDS with BOSS and 2dFLenS limited to the overlap region). 
Both results are consistent with our KiDS-1000 results, noting that the increase in our $S_8$ constraining power, by a factor of $\sim\! 2$ in this analysis, is driven by increases in both the KiDS survey area, and the analysed BOSS survey area.  

The impact of doubling the KiDS area can be seen by comparing to \citet{troester/etal:2020}, in Fig.~\ref{fig:DES_KiDS_comp}, who present a joint cosmic shear and galaxy clustering analysis of the KV450 KiDS release with the full BOSS area, finding $S_8 = 0.728 \pm 0.026$.   The $\sim\!40\%$ improvement in constraining power is consistent with expectations from the increased survey area, but a straightforward area-scaling comparison is inappropriate given that KiDS-1000 features improvements in the accuracy of the shear and photometric redshift calibrations, albeit at the expense of a decrease in the effective number density \citep[see][for details]{giblin/etal:inprep,hildebrandt/etal:inprep}.  

The offset in $S_8$ between the KiDS-1000-BOSS and KV450-BOSS $S_8$ constraints reflects a number of differences between the two analyses.  First, as the $S_8$ constraints from the \tttp analysis are primarily driven by KiDS (see Fig.~\ref{fig:cosmology-params-all}), we expect a reasonable statistical fluctuation in this parameter given the sampling variance arising from the significant increase in the KiDS survey area.  Using a simple model analysis in Appendix~\ref{app:expectedoffsets}, we conclude that we should expect differences, on average, of $|\Delta S_8| = 0.016$, and as such the increase that we find in $S_8$ between KV450 and KiDS-1000 is consistent with the expectation from simple statistical fluctuations.   BOSS primarily constrains $\Omega_{\rm m}$ which is impacted by the choice of prior on $n_{\rm s}$.  The wider $n_{\rm s}$ prior adopted in \citet{troester/etal:2020}, favours a slightly higher but less well-constrained value for $\Omega_{\rm m}$, leading to a slightly lower but less well-constrained value for $\sigma_8$, when combined with cosmic shear (see Appendix~\ref{app:priors}).   If we had also chosen an uninformative prior on $n_{\rm s}$ for our KiDS-1000-BOSS analysis, a decision that we cannot revise post unblinding, this would have likely served to exacerbate any tension with the {\it Planck} CMB constraints.

\subsection{Comparison with {\it Planck}}
\label{sec:planck_comp}
In our KiDS-1000-BOSS \tttp analysis we find good agreement with {\it Planck} for the matter density parameter, $\Omega_{\rm m}$, and the Hubble parameter, $h$, (see Fig.~\ref{fig:cosmology-params}).
The amplitude of matter fluctuations, $\sigma_8$, that we infer from the clustering of galaxies within, and lensing by, the large-scale structure of the low-redshift Universe is lower, however, than that inferred by {\it Planck}\footnote{A recent independent Atacama Cosmology Telescope CMB analysis reports $S_8=0.830 \pm 0.043$, in agreement with the {\it Planck} constraint of $S_8=0.834 \pm 0.016$ \citep[ACT,][]{aiola/etal:2020}.   Our results are fully consistent with the ACT CMB analysis, reflecting the larger uncertainty in the ACT constraints.} from the CMB. 

To quantify the level of discrepancy in the amplitude of matter fluctuations, we first concentrate on the parameter $S_{8} = \sigma_8 \sqrt{\Omega_{\rm m}/0.3}$ as it is tightly constrained and only exhibits negligible degeneracies, if at all, with the other cosmological parameters, $\Omega_{\rm m}$, $h$, and $n_{\rm s}$, as illustrated in Fig.~\ref{fig:cosmology-params-all}.   Comparing the reported marginal $S_8$ constraints, we find $S_8$ to be \kpoffperc lower than the CMB constraint from \citet{planck/etal:2018}.

We define the widely used $S_8$-difference measure
\begin{equation}
\label{eq: std_tension}
\tau = \frac{|\overline{S_{8}}^{\,\rm 3\times2pt}-\overline{S_{8}}^{\,\it Planck}|}{\sqrt{\mathrm{Var}[S_{8}^{\rm 3\times2pt}] + \mathrm{Var}[S_{8}^{\it Planck}]}} \,,
\end{equation}
where $\overline{S_{8}}$ and $\mathrm{Var}[S_{8}] $ denote the means and variances of the {\it Planck} and \tttp $S_8$ posterior distributions.  If both distributions are Gaussian, $\tau$ can be used to measure how likely it is that the mean of the difference between the distributions is consistent with zero.

Comparing the $S_{8}$ posterior distributions between our \tttp analysis and the {\it Planck} \software{plik\_lite\_TTTEEE}+\software{lowl}+\software{lowE} likelihood, we find $\tau=3.1$, meaning there is a $3.1\,\sigma$ difference between the KiDS-1000 and {\it Planck} constraints. 
Adopting two tension measures that do not assume Gaussianity of the marginal posterior distributions, both the `Hellinger' distance and the distribution of the $S_{8}$ parameter shifts indicate $3.1\,\sigma$ difference between our \tttp analysis and {\it Planck} (see Appendices~\ref{app:hellinger} and \ref{app:paramshift} for details).  
Our result thus continues the general trend of low-redshift probes preferring low amplitudes of matter fluctuations\footnote{Although we note the very recent clustering analysis released from e-BOSS which is fully consistent with Planck \citep{eBOSS/etal:2020}.} \citep{heymans/etal:2013, alam/etal:2017, leauthaud/etal:2017, abbott/etal:2018, hikage/etal:2019, bocquet/etal:2019, lange/etal:2019, palanque-delabrouille/etal:2020, wright/etal:2020b,DESclusters/etal:2020, singh/etal:2020}. 
In these cases the reported low $S_8$, or $\sigma_8$, constraints are formally statistically consistent with {\it Planck}, and well below the detection of any anomalies at the $5\,\sigma$-level. 
Considering, however, the $\sim\! 3\,\sigma$ difference that we have reported, and the overall trend in the literature, we would argue that we are reaching an uncomfortable point when it comes to regarding the $S_8$ offset as a simple statistical fluke.

\citet{Sanchez2020} pointed out that comparing $S_{8}$ between different experiments can be misleading due to the implicit dependence of $\sigma_{8}$ on $h$. 
Besides the intrinsic dependence of the amplitude of the matter power spectrum on $h$, measurements of $\sigma_{8}$ also depend on $h$ through the value of $8\,h^{-1}{\rm Mpc}$, the radius of the sphere within which the matter fluctuations are measured. 
In this way, constraints on $\sigma_8$ derived from data sets with different posterior distributions on $h$ represent the average of $\sigma(R)$ over different ranges of scales. 
We therefore also consider $S_{12} = \sigma_{12}\left(\Omega_{\rm m}\,h^2/0.14\right)^{0.4}$ \citep{Sanchez2020}, where $\sigma_{12}^{2}$ is the variance of the linear matter field at redshift zero in spheres of radius $12\,\mathrm{Mpc}$.
We find $S_{12} = 0.754^{+0.015}_{-0.018}$, with the value inferred by {\it Planck} being $S_{12} = 0.817_{-0.015}^{+0.011}
$, a difference of $\tau=3.0$, in agreement with the $S_8$ results. 

In light of the large parameter spaces that are being considered, we recognise that focussing on a single parameter can paint a simplistic picture of the agreement, or disagreement, between probes.
On a fundamental level, the question we wish to answer is whether a single model of the Universe can describe both the CMB as well as the low-redshift large-scale structure of the Universe.
Within our Bayesian inference framework, the Bayes factor provides a natural approach to model selection. 
The two models under consideration are 
\begin{description}
	\item[$\mathrm{M}_1$:] Both our \tttp data and {\it Planck}'s measurements of the CMB are described by a single flat \LCDM cosmology.
	\item[$\mathrm{M}_2$:] The two data sets are described by different cosmologies for the low- and high- redshift Universe, respectively.
\end{description}
The Bayes factor is then
\be
\label{equ:bayes-factor}
	R = \frac{P(\vec d | \mathrm{M}_1)P(\mathrm{M}_1)}{P(\vec d | \mathrm{M}_2)P(\mathrm{M}_2)} \ ,
\ee
where $P(\vec d | \mathrm{M}_i)$ is the probability of the data $\vec d$ under model $\mathrm{M}_i$ -- the Bayesian evidence. 

We assume the model priors $P(\mathrm{M}_1)$ and $P(\mathrm{M}_2)$ to be equal, that is, we make no a-priori assumption on the likelihood of $\mathrm{M}_1$ or $\mathrm{M}_2$. 
We use \software{anesthetic}\footnote{\url{https://github.com/williamjameshandley/anesthetic}}\citep{anesthetic} to compute $R$ and find $\ln R=3.1\pm0.3$, which can be interpreted as odds of $23\pm6$ in favour of model $\mathrm{M}_1$, and consistency between our \tttp measurement and {\it Planck}.  Given the dependence of $R$ on the parameter priors \citep{handley/lemos:2019}, we consider this result with some caution and review a series of alternative metrics that seek to quantify the tension between the full KiDS-1000 and {\it Planck} multi-dimensional cosmological parameter constraints.   These metrics are summarised in Table~\ref{tab:tension}, and below, with further details provided in Appendix~\ref{app:tensionest}.

\begin{table}
	\begin{center}
		\caption{Estimators of the consistency between our fiducial \tttp analysis and the \citet{planck/etal:2018} TTTEEE+lowE results. }
		\label{tab:tension}
\begin{tabular}{llccc}
    \toprule
   Metric   & Reference          & Value &PTE   & PTE [$\sigma$]\\
    \midrule
	$\tau$ & Eq.~(\ref{eq: std_tension})    & 3.1 & -- & \kpoff  \\
	$d_{\rm H}$ & Eq.~(\ref{eq:hellinger_def})& 0.95&-- & $3.1\,\sigma$\\
	$p_{\rm S}$ & Eq.~(\ref{equ:ps})& 0.9981& 0.0019&  $3.1\,\sigma$\\
    \midrule
	$R$   &  Eq.~(\ref{equ:bayes-factor}) & \kR & -- & -- \\
	$\ln R$  & Eq.~(\ref{eqn:logR}) &  \klogR& -- & --\\
	$\ln S$  &  Eq.~(\ref{equ:suspiciousness})  & \klogS & \klogSPTE &\klogSPTEsigma \\
	$Q_{\rm DMAP}$  & Eq.~(\ref{eqn:QDMAP})  & 9.5 & 0.037 & $2.1\, \sigma$\\
	$Q_{\rm UDM}$ &  Eq.~(\ref{equ:qudm})  & 7.7 & $0.054$  &$1.9\,\sigma$ \\
    \bottomrule
\end{tabular}
	\end{center}
	\tablefoot{The considered tension metrics are: 1) the differences between the means in $S_{8}$, divided by the standard deviations added in quadrature, $\tau$; 2) the Hellinger distance, $d_H$, between $S_8$ posterior distributions (Appendix~\ref{app:hellinger}); 3) the fraction of the $S_{8}$ parameter shift distribution that has a higher probability than no shifts (Appendix~\ref{app:paramshift}); 4) the Bayes ratio $R$ between a model that assumes a single cosmology for both our \tttp and the {\it Planck} data, and a model that uses separate cosmologies for the two data sets; 5) the logarithm $\ln R$ of the Bayes ratio; 6) the suspiciousness $\ln S$ \citep{handley/lemos:2019}; 7 \& 8) the concordance-discordance estimators $Q_{\rm DMAP}$ and $Q_{\rm UDM}$ \citep{Raveri2019}. 
		The columns list the estimator metric considered, the relevant equation reference, the measured value, and the `probability-to-exceed' the measured value (PTE), under the assumption that the two data sets are consistent.}
\end{table}

\citet{handley/lemos:2019} propose the `suspiciousness' statistic $S$ that is based on the Bayes factor, $R$, but hardened against prior dependences.   
We find that the probability of observing our measured suspiciousness statistic is $0.08\pm0.02$, which corresponds to a KiDS-{\it Planck} tension at the level of $1.8\pm0.1\,\sigma$ (see Appendix~\ref{app:sus} for details).

\citet{Raveri2019} introduce a number of metrics to quantify the consistency of data sets.   We consider their $Q_{\rm DMAP}$ metric, which explores the change in the goodness-of-fit when two data sets are combined.   A reduction in the goodness-of-fit can then be translated into a tension metric, finding a KiDS-{\it Planck} tension at the level of $2.1\sigma$ (see Appendix~\ref{app:QDMAP} for details).    
Their $Q_{\rm UDM}$ metric generalises the notion of parameter differences between posteriors to multiple dimensions, finding KiDS-{\it Planck} tension at the level of $1.9\,\sigma$ (see Appendix~\ref{app:QUDM} for details).   

Comparing tension metrics, we find a fairly consistent picture.  Reviewing tension in terms of the single parameter that our \tttp analysis is most sensitive to, $S_8$, leads to a $\sim\! 3\,\sigma$ tension with {\it Planck}.   
Including additional parameters into the tension analysis, parameters which KiDS is mainly insensitive to, serves to effectively dilute the tension, reducing the measure to the $\sim\! 2\,\sigma$ level.

\section{Conclusions}
\label{sec:conc}
In this analysis we have presented constraints on the flat $\Lambda$CDM cosmological model by combining observations of gravitational lensing and galaxy clustering to directly probe the evolution and distribution of the large-scale structures in the Universe.    Our survey of the $z \lesssim 1$ low-redshift Universe finds a matter distribution that is less clustered, compared to predictions from the best-fitting $\Lambda$CDM model to early-Universe CMB observations \citep{planck/etal:2018}.  This tendency for low-redshift probes to favour a smoother matter distribution compared to the CMB expectation has persisted since the first large-scale weak lensing survey \citep[CFHTLenS,][]{heymans/etal:2013}, but the significance of this effect has always been tantalisingly around, or below, the $\sim\! 3\,\sigma$ level.   It is therefore unclear if these differences are merely a statistical fluctuation, unaccounted for systematic errors, or a sign of interesting new physics.

Our new result does not lead to a resolution in the matter of statistical fluctuations, finding a \kpoff offset in the structure growth parameter $S_8 = \sigma_8 \sqrt{\Omega_{\rm m}/0.3}$ with $S_8=$\kSeightval.  Comparing the marginal $S_8$ constraints, we find $S_8$ to be \kpoffperc lower than the CMB constraint from \citet{planck/etal:2018}.   For a series of `tension' metrics that quantify differences in terms of the full posterior distributions, we find that the KiDS-1000 and {\it Planck} results agree at the $\sim\! 2\,\sigma$ level.   Through our series of image simulation analyses \citep{kannawadi/etal:2019}, catalogue null-tests \citep{giblin/etal:inprep}, variable depth mock galaxy survey analyses \citep{joachimi/etal:inprep}, optical-to-near-infrared photometric-spectroscopic redshift calibration, validated with mocks \citep{wright/etal:2020, vandenbusch/etal:2020,hildebrandt/etal:inprep}, internal consistency tests \citep[][Fig.~\ref{fig:cosmology-params-all} and Appendix~\ref{app:sensitivity}]{asgari/etal:inprep}, and marginalisation over a series of nuisance parameters that encompass our theoretical and calibration uncertainties (Appendix~\ref{app:priors}),  we argue that we have, however, addressed the question of \tttp systematic errors, robustly assessing and accounting for all sources of systematics that are known about in the literature.    

The KiDS-1000 cosmic shear constraints are highly complementary to the BOSS galaxy clustering constraints, leading to tight constraints in our joint \tttp analysis that are more than twice as constraining for the matter fluctuation amplitude parameter, $\sigma_8 = 0.760^{+0.021}_{-0.023}$, compared to previous \tttp analyses.    In the future, analysis of the clustering and galaxy-galaxy lensing of photometric samples with very accurate photometric redshifts \citep[see for example][]{vakili/etal:2019}, presents an opportunity for a future alternative KiDS-only \tttp photometric analysis, similar to the approach taken in \citet{abbott/etal:2018}.

In the next few years, two weak lensing surveys will see first light, with the launch of the {\it Euclid} satellite and the opening of the Vera~C.~Rubin Observatory.   These observatories will build the first two `full-sky' weak lensing surveys, which are highly complementary in terms of their differing strengths in depth and spatial resolution\footnote{The space-based {\it Nancy Grace Roman} Telescope is currently scheduled for launch in 2025 \citep{akeson/etal:2019}, joining {\it Euclid} and {\it Rubin} as an optimal weak lensing observatory for the future.}.  Combined with complementary overlapping redshift spectroscopy from DESI, 4MOST and {\it Euclid}, the multi-probe weak lensing and spectroscopic galaxy clustering methodology, which we have implemented in this analysis, provides a promising route forward for these next generation surveys.   We view this \tttp approach as just the start of the story, however, looking forward to a future combined analysis of weak lensing and galaxy clustering with both photometric and spectroscopic lenses, a combination which we call a `$6\times2$pt' approach \citep{bernstein:2009}.    This would allow for the optimal combination of information from the clustering cross-correlation of spectroscopic and photometric galaxies \citep{newman:2008}, an observable that we currently only use as an independent tool to validate our photometric redshift calibration \citep{hildebrandt/etal:inprep}.      Developments in the area of highly non-linear galaxy bias, baryon feedback and intrinsic alignment modelling, along with a sufficiently flexible but tractable redshift distribution model and an accurate `$6\times2$pt' covariance estimate, will all be required in order to realise this long-term goal.   The effort will, however, be worthwhile allowing for the implementation of arguably the most robust methodology available to mitigate systematic errors, whilst simultaneously enhancing cosmological parameter constraints.

The ESO-KiDS public survey completed observations in July 2019, spanning $1350\,\mathrm{deg}^{2}$.   We therefore look forward to the fifth and final KiDS data release, `KiDS-Legacy', along with new results from the concurrent `Stage-III' surveys, DES and HSC, whilst the community prepares for the next exciting chapter of `full-sky' weak lensing surveys.  

\begin{acknowledgements}
We are completely indebted to Eric Tittley at the IfA for going well beyond the call of duty to save the KiDS-1000 data products after an explosion in our server room destroyed the RAID.   We thank our anonymous referee for their positive feedback and useful comments, in addition to our external blinder Matthias Bartelmann who revealed the key for which of the three catalogues analysed was the true unblinded catalogue on the 9th July 2020, right at the end of the KiDS-1000 study which was submitted to A\&A on the 30th July 2020.   We also wish to thank the Vera C. Rubin Observatory LSST-DESC Software Review Policy Committee (Camille Avestruz, Matt Becker, Celine Combet, Mike Jarvis, David Kirkby, Joe Zuntz with CH) for their draft Software Policy document which we followed, to the best of our abilities, during the KiDS-1000 project.   Following this draft policy, the software used to carry out the various analyses presented in this paper is open source at \href{https://github.com/KiDS-WL/Cat_to_Obs_K1000_P1}{github.com/KiDS-WL/Cat\textunderscore to\textunderscore Obs\textunderscore K1000\textunderscore P1} and \href{https://github.com/KiDS-WL/KCAP}{github.com/KiDS-WL/KCAP}. 
The figures in this work were created with \software{matplotlib} \citep{Hunter2007} and \software{getdist} \citep{Lewis2019}, making use of the 
\software{numpy} \citep{Oliphant2006} and \software{scipy} \citep{Jones2001} software packages. We also made extensive use of the {\sc TreeCorr} and {\sc CosmoSIS} software packages and thank
Mike Jarvis and Joe Zuntz for their continuing enhancements and clear documentation.\\

This project has received significant funding from the European Union's Horizon 2020 research and innovation programme.  We thank and acknowledge support from: the European Research Council under grant agreement No.~647112 (CH, TT, MA, CL and BG), No.~770935 (HHi, AHW, JLvdB and AD) and No.~693024 (SJ) in addition to the Marie Sk\l{}odowska-Curie grant agreement No.~797794 (TT).   We also acknowledge support from the Max Planck Society and the Alexander von Humboldt Foundation in the framework of the Max Planck-Humboldt Research Award endowed by the Federal Ministry of Education and Research (CH, FK);  the Deutsche Forschungsgemeinschaft Heisenberg grant Hi 1495/5-1, (HHi);  the Netherlands Organisation for Scientific Research Vici grant 639.043.512 (AK, HHo) and grant 621.016.402 (JdJ); the Alexander von Humboldt Foundation (KK);  the Polish Ministry of Science and Higher Education through grant DIR/WK/2018/12, and the Polish National Science Center through grants no. 2018/30/E/ST9/00698 and 2018/31/G/ST9/03388 (MB); the Royal Society through an Enhancement Award RGF/EA/181006 (BG);  the Australian Research Council grants DP160102235 and CE17010013 (KG);  the Beecroft Trust (SJ);  STFC grant ST/N000919/1 (LM);  the Netherlands Research School for Astronomy and Target (GVK); the NSFC of China under grant 11973070, the Shanghai Committee of Science and Technology grant No.19ZR1466600, and the Key Research Programme of Frontier Sciences, CAS, Grant No. ZDBS-LY-7013 (HYS).\\

The results in this paper are based on observations made with ESO Telescopes at the La Silla Paranal Observatory under programme IDs 177.A-3016, 177.A-3017, 177.A-3018 and 179.A-2004, and on data products produced by the KiDS consortium. The KiDS production team acknowledges support from: Deutsche Forschungsgemeinschaft, ERC, NOVA and NWO-M grants; Target; the University of Padova, and the University Federico II (Naples).  Data processing for VIKING has been contributed by the VISTA Data Flow System at CASU, Cambridge and WFAU, Edinburgh. \\

The BOSS-related results in this paper have been made possible thanks to SDSS-III. Funding for SDSS-III has been provided by the Alfred P. Sloan Foundation, the Participating Institutions, the National Science Foundation, and the U.S. Department of Energy Office of Science.   SDSS-III is managed by the Astrophysical Research Consortium for the Participating Institutions of the SDSS-III Collaboration including the University of Arizona, the Brazilian Participation Group, Brookhaven National Laboratory, Carnegie Mellon University, University of Florida, the French Participation Group, the German Participation Group, Harvard University, the Instituto de Astrofisica de Canarias, the Michigan State/Notre Dame/JINA Participation Group, Johns Hopkins University, Lawrence Berkeley National Laboratory, Max Planck Institute for Astrophysics, Max Planck Institute for Extraterrestrial Physics, New Mexico State University, New York University, Ohio State University, Pennsylvania State University, University of Portsmouth, Princeton University, the Spanish Participation Group, University of Tokyo, University of Utah, Vanderbilt University, University of Virginia, University of Washington, and Yale University.\\

The 2dFLenS-related results are based on data acquired through the Australian Astronomical Observatory, under programme A/2014B/008. It would not have been possible without the dedicated work of the staff of the AAO in the development and support of the 2dF-AAOmega system, and the running of the AAT.\\

{ {\it Author contributions:}  All authors contributed to the development and writing of this paper.  The authorship list is given in three groups:  the lead authors (CH \& TT) followed by two alphabetical groups.  The first alphabetical group includes those who are key contributors to both the scientific analysis and the data products.  The second group covers those who have either made a significant contribution to the data products, or to the scientific analysis.}
\end{acknowledgements}

\bibliographystyle{aa} 
\bibliography{references} 

\begin{appendix} 
\section{Galaxy properties and the \tttp covariance}
\label{app:properties}

In this appendix we tabulate the properties of the KiDS-1000 tomographic source samples, along with the properties of the BOSS and 2dFLenS lens samples, in Table~\ref{tab:datatab}.   
We list the spectroscopic redshift selection for the lenses ($z_{\rm min} < z_{\rm s} \leq z_{\rm max}$), and the photometric redshift selection for the sources ($z_{\rm min} < z_{\rm B} \leq z_{\rm max}$), along with the mean redshift of each sample.  
For the source sample the true redshift distributions are estimated in \citet{hildebrandt/etal:inprep}, using the SOM methodology from \citet{wright/etal:2020}.     
The shear calibration correction, $m$, which can also be referred to in the literature as the responsivity, $R = 1+m$, is listed for each source bin \citep{kannawadi/etal:2019}.  
The effective number density of lenses and sources defines the number of galaxies per square arcminute in the case of unit weights and, for the sources, unit responsivity \citep[see equations C.11 and C.13 in][]{joachimi/etal:inprep}.  
We also list the effective ellipticity dispersion, $\sigma_{\epsilon,i}$, per ellipticity component, $i$, for each the weighted and calibrated source galaxy samples \citep[equation C.8 in][]{joachimi/etal:inprep}.

\begin{table}
\caption{Galaxy properties for the BOSS and 2dFLenS lens (\lq L\rq) samples and the KiDS-1000 source (\lq S\rq) samples.}              
\label{tab:datatab}      
\centering                                      
\begin{tabular}{lcccccr}          
\toprule
ID & $z_{\rm min}$ &  $z_{\rm max}$& mean $z$ & $n_{\rm eff}$ & $\sigma_{\epsilon,i}$ & \multicolumn{1}{c}{$m$}\\    
\midrule
\multicolumn{6}{l}{\bf KiDS-1000:}\\  
S1 & 0.1 & 0.3 & 0.26 & 0.62 &  0.27 & $-0.009\pm0.019$\\
S2 & 0.3 & 0.5 & 0.40 & 1.18 &  0.26 & $-0.011\pm0.020$\\
S3 & 0.5 & 0.7 & 0.56 & 1.85 &  0.27 & $-0.015\pm0.017$\\
S4 & 0.7 & 0.9 & 0.79 & 1.26 &  0.25 & $0.002\pm0.012$\\
S5 & 0.9 & 1.2 & 0.98 & 1.31 &  0.27 & $0.007\pm0.010$\\
\midrule      
\multicolumn{6}{l}{\bf BOSS:}\\                             
L1 & 0.2 & 0.5 & 0.38 & $0.014$ & -  & \multicolumn{1}{c}{-}\\
L2 & 0.5 & 0.75 & 0.61 & $0.016$ & -  & \multicolumn{1}{c}{-}\\
\midrule      
\multicolumn{6}{l}{\bf 2dFLenS:}\\                                
L1 & 0.2 & 0.5 & 0.36 & $0.006$ & - & \multicolumn{1}{c}{-}\\
L2 & 0.5 & 0.75 & 0.60 & $0.006$ & - & \multicolumn{1}{c}{-}\\
\bottomrule
\end{tabular}
\tablefoot{Columns include the bin identifier, ID, the minimum and maximum redshift selection for the bin, $z_{\rm min/max}$, which applies to photometric redshifts for the sources S, and spectroscopic redshifts for the lenses L, along with the mean redshift of the bin, and the effective galaxy number density, per square arcminute, $n_{\rm eff}$.  For the source bins we also include the measured ellipticity dispersion per component, $\sigma_{\epsilon,i}$, and the shear calibration correction, $m$, and its uncertainty.}
\end{table}

\begin{figure}
	\begin{center}
		\includegraphics[width=\columnwidth]{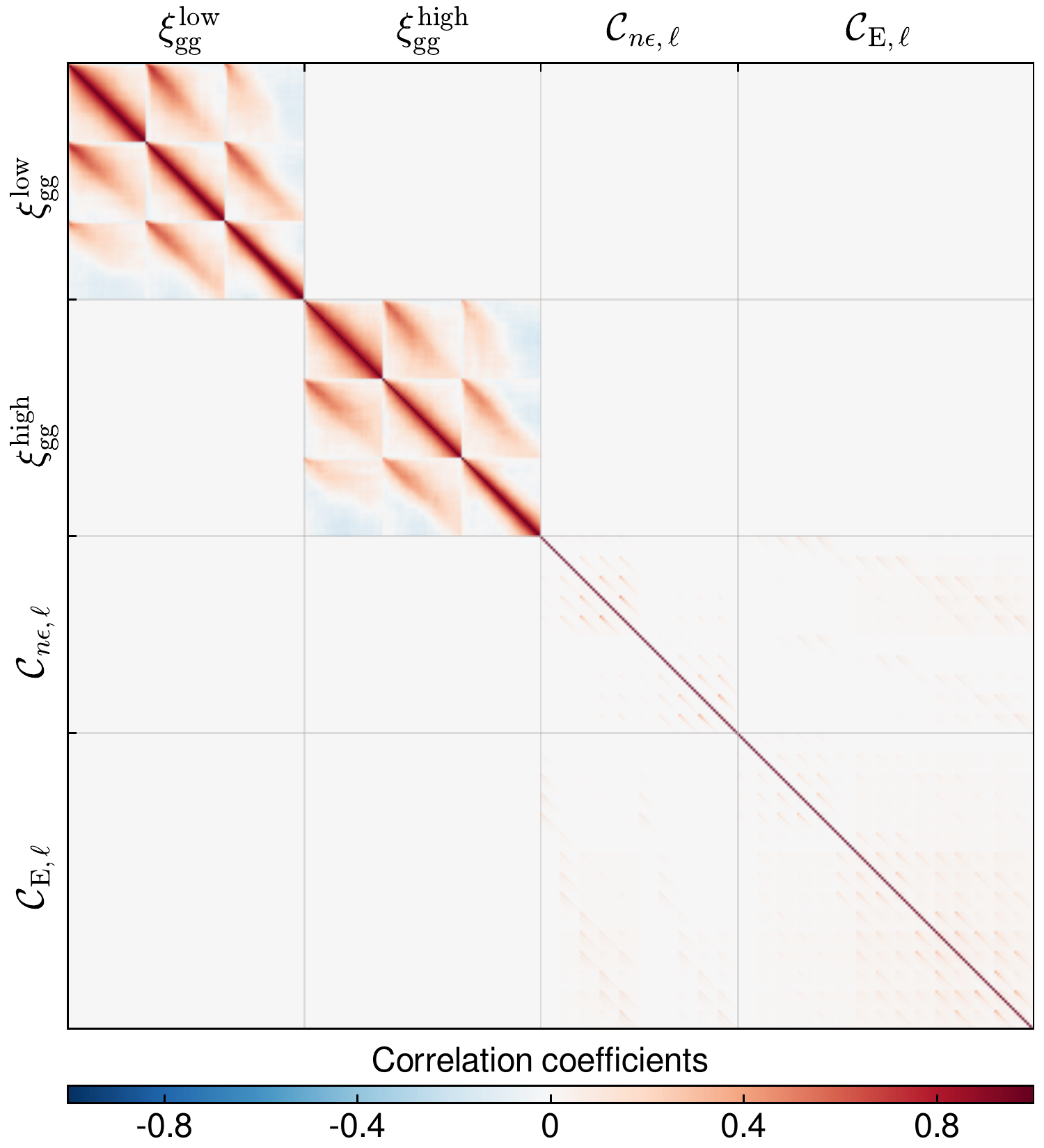}
		\caption{Correlation coefficients of the \tttp covariance matrix for the cosmic shear, $\mathcal{C}_{\rm E}$, galaxy-galaxy lensing, $\mathcal{C}_{n\epsilon}$ and galaxy clustering observables, $\xi_{\rm gg}$ (see Sect.~\ref{sec:data} for details).  Here the band powers $\mathcal{C}$, are related to the angular power spectrum $C$, in Eq.~(\ref{eq:cl_cosmicshear}) and Eq.~(\ref{eq:pgg}), as $\mathcal{C}= \ell^2 C/(2\pi)$.}
		\label{fig:ttttpcov}
	\end{center}
\end{figure}

Fig.~\ref{fig:ttttpcov} displays the correlation coefficients of the \tttp covariance matrix for the three observables; cosmic shear E-mode power spectra, $\mathcal{C}_E$, galaxy-galaxy lensing E-mode power spectra, $\mathcal{C}_{n\epsilon}$, and the anisotropic galaxy clustering in low and high redshift bins, $\xi_{\rm gg}$ (see Sect.~\ref{sec:data} for details).   The cross-correlation between the two lensing and the clustering observables is set to zero, as mock data analyses showed these correlations to be negligible for the KiDS and BOSS footprints \citep{joachimi/etal:inprep}.  The Fourier-space lensing observables are shown to be significantly less correlated between $\ell$-scales, in comparison to the physical-scale clustering observables.

\section{Parameter priors}
\label{app:priors}
In this appendix we tabulate the adopted KiDS-1000 priors and sampling parameters in Table~\ref{tab:priors}.   
The uniform prior on the dimensionless Hubble constant, $h$, reflects distance-ladder $\pm 5\,\sigma$ constraints from \citet{riess/etal:2016}, which encompasses the value of $h$ favoured by \citet{planck/etal:2018}.  
The uniform prior on the baryon density, $\omega_{\rm b}= \Omega_{\rm b}h^2$, reflects big bang nucleosynthesis $\pm 5\, \sigma$ constraints from \citet{olive/etal:2014}.   
The uniform prior on the CDM density, $\omega_{\rm c} = \Omega_{\rm c}h^2$, reflects Supernova Type Ia $\pm 5\, \sigma$ constraints on $\Omega_{\rm m}$ from \citet{scolnic/etal:2018} combined with the most extreme allowed values of $h$ and $\omega_{\rm b}$, given their priors.   

As discussed in Sect.~\ref{sec:KCAP} we choose to sample with an uninformative uniform prior on $S_8$ to avoid implicit informative priors from a uniform prior on the primordial power spectrum amplitude $A_{\rm s}$.    
We choose a fixed model for the properties of neutrinos, adopting the normal hierarchy at the minimum sum of masses, $\Sigma m_\nu = 0.06\,\mathrm{eV}$, following \citet{planck/etal:2018}.  
We consider extended models, including variations on neutrino mass in \citet{troester/etal:inprep}.

\begin{table}
\caption{KiDS-1000 sampling parameters and priors.}              
\label{tab:priors}      
\centering                                      
\begin{tabular}{lll}          
\toprule
Parameter & Symbol & Prior \\    
\midrule                                   
Hubble constant & $h$ & $\bb{0.64,\,0.82}$ \\
Baryon density & $\omega_{\rm b}$ & $\bb{0.019,\,0.026}$ \\
CDM density & $\omega_{\rm c}$ & $\bb{0.051,\,0.255}$ \\
Density fluctuation amp. & $S_8$ & $\bb{0.1,\,1.3}$ \\
Scalar spectral index & $n_{\rm s}$ & $\bb{0.84,\,1.1}$ \\
\midrule
Linear galaxy bias & $b_1 \;[2]$ & $\bb{0.5,\,9}$ \\
Quadratic galaxy bias & $b_2 \;[2]$ & $\bb{-4,\,8}$ \\
Non-local galaxy bias & ${\gamma_3^-} \;[2]$ & $\bb{-8,\,8}$ \\
Virial velocity parameter & $a_{\rm vir} \;[2]$ & $\bb{0,\,12}$ \\
Intrinsic alignment amp. & $A_{\rm IA}$ & $\bb{-6,\,6}$ \\
Baryon feedback amp. & $A_{\rm bary}$ & $\bb{2,\,3.13}$ \\
\midrule
Redshift offsets & ${\bf \delta_z}$ & ${\cal N}(\vek{\mu};\vek{C}_{\delta z})$ \\
\bottomrule
\end{tabular}
\tablefoot{Primary cosmological parameters for the flat $\Lambda$CDM model are listed in the first section. The second section lists astrophysical nuisance parameters to model galaxy bias (with independent parameters for each of the two BOSS redshift bins as indicated with the bracket $[2]$), intrinsic galaxy alignments, and baryon feedback.  Observational redshift nuisance parameters are listed in the final section. Prior values in square brackets are the limits of the adopted uniform top-hat priors.  ${\cal N}(\mu;C)$ corresponds to a five dimensional multivariate Gaussian prior with mean, $\vek{\mu}$, and covariance, $\vek{C}_{\delta z}$.}
\end{table}

\begin{figure}
	\begin{center}
		\includegraphics[width=\columnwidth]{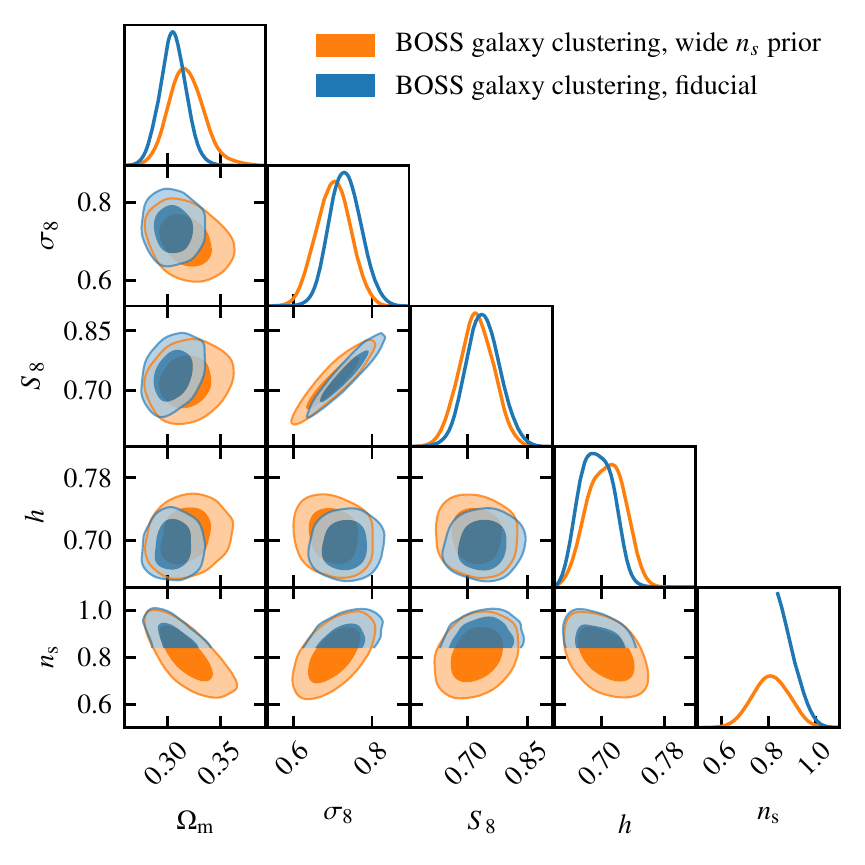}
		\caption{Impact of the $n_{\rm s}$ prior: Comparing marginalised posterior distributions for the BOSS galaxy clustering analysis for our fiducial analysis (blue) with the constraints when adopting an uninformative prior on $n_{\rm s}$ (orange).   Opening the parameter space to arguably unphysical values of $n_{\rm s}$ favours higher values and weaker constraints on $\Omega_{\rm m}$. Constraints on $S_8$ are, however, fairly insensitive to the choice of $n_{\rm s}$ prior.}
		\label{fig:ns-prior}
	\end{center}
\end{figure}

The uniform prior on the scalar spectral index, $n_{\rm s}$, reflects a restriction in our likelihood implementation, where the \citet{sanchez/etal:2017} galaxy clustering likelihood becomes prohibitively slow for $n_{\rm s}>1.1$.  With the upper limit of the top-hat prior fixed by this computational limitation, we choose to symmetrise the prior around the theoretical expectation of $n_{\rm s}=0.97$.     In Fig.~\ref{fig:ns-prior} we demonstrate the impact of this informative prior on the BOSS-only galaxy clustering constraints, highlighting how informative this choice of prior is.   We argue that adopting an informative prior is justified however, given our theoretical prior knowledge of the Harrison-Zel'dovich spectrum.   Fig.~\ref{fig:ns-prior} also helps to illustrate that had we chosen an uninformative prior on $n_{\rm s}$ for our KiDS-1000-BOSS analysis, a decision taken more than a year before unblinding our analysis, this would have likely served to exacerbate any tension with the Planck CMB constraints. 

Turning to astrophysical priors, the galaxy bias parameter top-hat priors on $b_1$, $b_2$,  $\gamma_3^-$, (Eq.~\ref{eq:pgg}), and on $a_{\rm vir}$ \citep[see the `fingers of god' model in equations 6 to 9 of][]{joachimi/etal:inprep},
match those adopted in \citet{troester/etal:2020}, which cover a wider range than those used in \citet{sanchez/etal:2017}.  
The two BOSS redshift slices have independent sets of parameters.   
Wide uniform priors for the intrinsic alignment parameter $A_{\rm IA}$ are chosen to be uninformative.    
Uniform priors on the baryon feedback parameter $A_{\rm bary}$ are chosen such that the resulting \citet{mead/etal:2016} model of the non-linear matter power spectrum encompasses both the most aggressive feedback model from the \citet{vandaalen/etal:2011} suite of hydrodynamical simulations, along with the dark matter-only case where $A_{\rm bary}=3.13$.

There are five additional correlated nuisance parameters, $\delta^i_z$, that model uncertainty in the mean of the source redshift distributions.  We adopt a multivariate Gaussian prior for the vector $\vek{\delta}_z$ with a mean $\vek{\mu} = (0.0001,0.0021,0.0129,0.0110,-0.0060)$, and a covariance, $C_{\delta z}$, as calibrated using mock galaxy catalogues in \citet{wright/etal:2020}.   
The diagonal terms of $\vek{C}_{\delta z}$ are typically at the level of $\sim\!(0.01)^2$, with off-diagonal correlation coefficients ranging between $\sim\! 0.1$ and $\sim\! 0.3$ \citep[see section 3 and figure 2 of][for details]{hildebrandt/etal:inprep}.

\section{Parameter constraints}
\label{app:parameter-constraints}
In this appendix we tabulate the maximum posterior (MAP) and marginalised constraints on the flat $\Lambda$CDM cosmological parameters, in Table~\ref{tab:fullparams}, for the different combinations of the three large-scale structure probes considered in this work.   For constraints from KiDS-1000 cosmic shear alone, we refer the reader to \citet{asgari/etal:inprep}.
\begin{landscape}
\begin{table}
\begin{center}
\caption{Parameter constraints for the probe combinations considered in this work: \tttp, cosmic shear (CS) and galaxy-galaxy lensing (GGL), cosmic shear and galaxy clustering (GC), and galaxy clustering by itself. 
We refer the reader to \citet{asgari/etal:inprep} for cosmic shear-only constraints. 
For each probe, the first column lists the MAP value with the PJ-HPD CI. 
If the PJ-HPD CI could not be robustly determined, no uncertainty estimate is provided. 
The second column for each probe lists the maximum of the marginal posterior, together with the marginal HPD CI. 
Parameter for which the marginal probability at either prior edge exceeds 13\% of the peak probability are deemed unconstrained \citep[see appendix A of][]{asgari/etal:inprep}, and are denoted by a dash. 
Finally, parameters that are not sampled are left blank.
\label{tab:fullparams}}
\begin{tabular}{lllllllll}
    \toprule
    Parameter    & $3\times2$pt & $3\times2$pt& GC & GC& CS+GGL & CS+GGL& CS+GC & CS+GC \\ 
             & (joint) & (marginal)& (joint) & (marginal)& (joint) & (marginal)& (joint) & (marginal) \\ 

    \midrule
$S_8     $& $0.766^{+0.02}_{-0.014}$ & $0.765^{+0.017}_{-0.016}$& $0.762^{+0.054}_{-0.05}$ & $0.743^{+0.038}_{-0.05}$& $0.774^{+0.031}_{-0.029}$ & $0.761^{+0.024}_{-0.025}$& $0.766^{+0.018}_{-0.015}$ & $0.763^{+0.016}_{-0.016}$\\ [0.3 em]
$h^2\Omega_\mathrm{c}$& $0.123$ & $0.124^{+0.0096}_{-0.011}$& $0.136^{+0.0065}_{-0.018}$ & $0.123^{+0.011}_{-0.009}$& $0.0721^{+0.089}_{-0.016}$ & $0.12^{+0.057}_{-0.034}$& $0.126^{+0.0075}_{-0.013}$ & $0.124^{+0.01}_{-0.011}$\\ [0.3 em]
$h^2\Omega_\mathrm{b}$& $0.0225$ & --& $0.0246^{+0.0012}_{-0.0037}$ & --& $0.0258$ & --& $0.0219^{+0.0022}_{-0.0022}$ & --\\ [0.3 em]
$h       $& $0.695^{+0.03}_{-0.019}$ & $0.696^{+0.02}_{-0.026}$& $0.716^{+0.02}_{-0.035}$ & $0.687^{+0.029}_{-0.018}$& $0.642^{+0.1}_{-1.2e-05}$ & --& $0.695^{+0.022}_{-0.023}$ & $0.69^{+0.026}_{-0.02}$\\ [0.3 em]
$n_\mathrm{s}$& $0.901$ & --& $0.851^{+0.06}_{-0.01}$ & --& $0.947$ & --& $0.894^{+0.05}_{-0.042}$ & --\\ [0.3 em]
\midrule
$A_{\rm bary}$& $3.13^{+1.6e-05}_{-0.47}$ & --& $$ & & $3.13$ & --& $3.13$ & --\\ [0.3 em]
$A_{\rm IA}$& $1.07^{+0.27}_{-0.31}$ & $1.01^{+0.31}_{-0.28}$& $$ & & $0.974^{+0.32}_{-0.2}$ & $0.972^{+0.25}_{-0.26}$& $0.947^{+0.44}_{-0.31}$ & $0.892^{+0.35}_{-0.36}$\\ [0.3 em]
$\delta \bar{z_1}$& $1.93^{+11}_{-10}\times 10^{-3}$ & $2.8^{+8.8}_{-12}\times 10^{-3}$& $$ & & $1.41^{+8.4}_{-11}\times 10^{-3}$ & $2.15^{+8.7}_{-11}\times 10^{-3}$& $2.02^{+7.4}_{-14}\times 10^{-3}$ & $2.45^{+9.5}_{-12}\times 10^{-3}$\\ [0.3 em]
$\delta \bar{z_2}$& $0.0101^{+0.016}_{-0.0068}$ & $9.09^{+11}_{-9.7}\times 10^{-3}$& $$ & & $0.0102^{+0.014}_{-0.0069}$ & $8.75^{+11}_{-8.7}\times 10^{-3}$& $0.0117^{+0.013}_{-0.0085}$ & $0.0121^{+0.0095}_{-0.012}$\\ [0.3 em]
$\delta \bar{z_3}$& $-0.0207^{+0.0093}_{-0.01}$ & $-0.0204^{+0.01}_{-0.0093}$& $$ & & $-0.0206^{+0.0091}_{-0.0091}$ & $-0.0199^{+0.0092}_{-0.0092}$& $-0.013^{+0.0091}_{-0.012}$ & $-0.0119^{+0.0098}_{-0.011}$\\ [0.3 em]
$\delta \bar{z_4}$& $-0.0143^{+0.0076}_{-0.008}$ & $-0.0145^{+0.0082}_{-0.0076}$& $$ & & $-0.0138^{+0.0066}_{-0.0086}$ & $-0.012^{+0.0062}_{-0.009}$& $-0.0166^{+0.0098}_{-0.0064}$ & $-0.0156^{+0.0077}_{-0.0085}$\\ [0.3 em]
$\delta \bar{z_5}$& $5.23^{+8.3}_{-10}\times 10^{-3}$ & $4.54^{+10}_{-8.7}\times 10^{-3}$& $$ & & $6.06^{+7.3}_{-10}\times 10^{-3}$ & $6.85^{+8.7}_{-8.6}\times 10^{-3}$& $6.5^{+12}_{-7.1}\times 10^{-3}$ & $6.54^{+9.4}_{-9}\times 10^{-3}$\\ [0.3 em]
$b_1^{\rm lowz}$& $2^{+0.079}_{-0.083}$ & $2.04^{+0.064}_{-0.093}$& $2.01^{+0.13}_{-0.16}$ & $2.09^{+0.11}_{-0.13}$& $3.29^{+0.3}_{-0.81}$ & $3.14^{+0.55}_{-0.55}$& $1.99^{+0.09}_{-0.07}$ & $2.02^{+0.085}_{-0.07}$\\ [0.3 em]
$b_2^{\rm lowz}$& $0.3^{+0.62}_{-0.69}$ & $0.289^{+0.71}_{-0.61}$& $0.475^{+1.1}_{-0.83}$ & $0.624^{+0.96}_{-1}$& $0.968$ & --& $0.443^{+1}_{-0.67}$ & $0.326^{+1}_{-0.67}$\\ [0.3 em]
$\gamma_{3-}^{\rm lowz}$& $0.695^{+0.48}_{-0.54}$ & $0.879^{+0.44}_{-0.52}$& $0.688^{+0.59}_{-0.57}$ & $0.752^{+0.62}_{-0.56}$& $7.87$ & --& $0.594^{+0.58}_{-0.57}$ & $0.853^{+0.53}_{-0.52}$\\ [0.3 em]
$a_{\rm vir}^{\rm lowz}$& $4.14^{+0.84}_{-0.97}$ & $4.24^{+0.86}_{-0.96}$& $4.3^{+1.1}_{-1.2}$ & $4.44^{+1.1}_{-1.2}$& $$ & & $4.28^{+1.2}_{-1}$ & $4.37^{+1.1}_{-1.1}$\\ [0.3 em]
$b_1^{\rm highz}$& $2.13^{+0.095}_{-0.091}$ & $2.17^{+0.083}_{-0.093}$& $2.14^{+0.16}_{-0.16}$ & $2.23^{+0.13}_{-0.14}$& $2.3^{+0.33}_{-0.68}$ & $2.26^{+0.48}_{-0.52}$& $2.12^{+0.1}_{-0.094}$ & $2.17^{+0.092}_{-0.084}$\\ [0.3 em]
$b_2^{\rm highz}$& $-1.15^{+0.75}_{-0.41}$ & $-1.03^{+0.72}_{-0.46}$& $-0.948^{+2.3}_{-0.75}$ & $-0.901^{+2.3}_{-1}$& $5.82^{+2}_{-3.2}$ & --& $-1.16^{+1.8}_{-0.48}$ & $-0.951^{+1.7}_{-0.85}$\\ [0.3 em]
$\gamma_{3-}^{\rm highz}$& $0.369^{+0.8}_{-0.79}$ & $0.505^{+0.9}_{-0.62}$& $-0.126^{+0.94}_{-1.1}$ & $0.452^{+0.7}_{-1.2}$& $8^{+0.00058}_{-5.4}$ & --& $-0.0141^{+0.91}_{-0.91}$ & $0.412^{+0.8}_{-0.81}$\\ [0.3 em]
$a_{\rm vir}^{\rm highz}$& $1.18^{+1.5}_{-0.97}$ & --& $1.84^{+2.4}_{-1.8}$ & --& $$ & & $1.32^{+2.7}_{-1.3}$ & --\\ [0.3 em]
\midrule
$\Omega_\mathrm{m}$& $0.305^{+0.01}_{-0.015}$ & $0.306^{+0.012}_{-0.013}$& $0.313^{+0.0097}_{-0.017}$ & $0.307^{+0.011}_{-0.015}$& $0.239^{+0.11}_{-0.076}$ & $0.29^{+0.087}_{-0.07}$& $0.306^{+0.013}_{-0.012}$ & $0.307^{+0.012}_{-0.014}$\\ [0.3 em]
$\sigma_8$& $0.76^{+0.025}_{-0.02}$ & $0.759^{+0.02}_{-0.024}$& $0.75^{+0.045}_{-0.047}$ & $0.725^{+0.048}_{-0.036}$& $0.867^{+0.2}_{-0.17}$ & $0.729^{+0.13}_{-0.098}$& $0.758^{+0.028}_{-0.018}$ & $0.758^{+0.019}_{-0.024}$\\ [0.3 em]
$\sigma_{12}$& $0.743^{+0.03}_{-0.026}$ & $0.737^{+0.029}_{-0.027}$& $0.717^{+0.058}_{-0.035}$ & $0.707^{+0.046}_{-0.04}$& $0.887^{+0.15}_{-0.23}$ & $0.72^{+0.095}_{-0.14}$& $0.734$ & $0.732^{+0.03}_{-0.024}$\\ [0.3 em]
$S_{12}  $& $0.754^{+0.021}_{-0.013}$ & $0.754^{+0.015}_{-0.018}$& $0.751^{+0.038}_{-0.054}$ & $0.728^{+0.041}_{-0.045}$& $0.77^{+0.067}_{-0.046}$ & $0.757^{+0.036}_{-0.05}$& $0.753^{+0.022}_{-0.013}$ & $0.754^{+0.014}_{-0.019}$\\ [0.3 em]
$A_\mathrm{s}$& $1.85^{+0.25}_{-0.17}\times 10^{-9}$ & $1.79^{+0.22}_{-0.18}\times 10^{-9}$& $1.7^{+0.33}_{-0.16}\times 10^{-9}$ & $1.64^{+0.27}_{-0.18}\times 10^{-9}$& $5.34^{+5.2}_{-4.3}\times 10^{-9}$ & --& $1.81^{+0.27}_{-0.15}\times 10^{-9}$ & $1.78^{+0.21}_{-0.18}\times 10^{-9}$\\ [0.3 em]
$100\theta_{\rm MC}$& $1.05$ & $1.05^{+0.011}_{-0.015}$& $1.06^{+0.0082}_{-0.019}$ & $1.05^{+0.012}_{-0.013}$& $0.96^{+0.13}_{-0.023}$ & $1.09^{+0.039}_{-0.058}$& $1.05^{+0.009}_{-0.016}$ & $1.05^{+0.012}_{-0.014}$\\ [0.3 em]
    \bottomrule
\end{tabular}

\end{center}
\end{table}
\end{landscape}

\section{Modelling intrinsic galaxy alignment}
\label{app:IAmodel}
In this analysis we adopt the \citet{bridle/king:2007} NLA model in order to marginalise over our uncertainty in the contribution to the observed two-point shear correlation function from the intrinsic alignment (IA) of galaxies within their surrounding density field.   More sophisticated models exist, however, and in this appendix we briefly discuss these alternatives.   We then provide justification for our choice by summarising the analysis of \citet{fortuna/etal:2020} which demonstrates that for the statistical power of KiDS-1000, the use of the somewhat adhoc NLA model is sufficiently flexible so as not to introduce any biases in the cosmological parameter constraints.

There are two advanced methods to determine a non-linear IA model.    The first uses a perturbative approach to model the non-linear behaviour of the tidal alignment (where the galaxy is preferentially aligned with the `stretching axis' of the tidal quadrupole) and tidal torquing (where the galaxy disc forms perpendicular to the angular momentum axis which is dependent on the tidal field) \citep{blazek/etal:2019, vlah/etal:2020}.  The second uses a halo model approach, which introduces a model for the small-scale alignment of satellite galaxies within central haloes \citep{schneider/bridle:2010}, and allows for different alignment strengths to be included for the evolving red and blue galaxy population \citep{fortuna/etal:2020}.  On large physical scales both techniques, along with the NLA model, recover the linear alignment model of \citet{hirata/seljak:2004}.  On small physical scales the accuracy of the perturbative approach is limited by the order of the corrections adopted \citep{blas/etal:2013}.  The accuracy of the halo model approach is limited by the challenge of modelling the transition between the properties of galaxies within single halos and across the full density field, commonly referred to as the one-to-two halo transition \citep[see for example the discussion in][]{mead/etal:2020b}.   Both models are nevertheless a significant improvement on the NLA model, which developed an adhoc solution to a small-scale mismatch of the \citet{hirata/seljak:2004} model with IA numerical simulations \citep{heymans/etal:2004}, by replacing the linear matter power spectrum in the \citet{hirata/seljak:2004} formalism with the non-linear matter power spectrum.

The IA observations of red and blue galaxies have been both direct \citep{joachimi/etal:2011,mandelbaum/etal:2011,singh/etal:2015,tonegawa/etal:2018,johnston/etal:2019} and indirect \citep{heymans/etal:2013, samuroff/etal:2019}, recovering a common conclusion of significant alignment for red galaxies, and, as yet, no detection of alignment for blue galaxies.   Direct observations are limited by the depth of the spectroscopic, or high-accuracy photometric redshift, galaxy samples that can be studied.  Indirect observations, where an IA model is constrained simultaneously with a cosmological model, are limited by degeneracies with nuisance parameters.   As the IA and cosmological signal scale differently with redshift, a flexible IA model can absorb any systematic errors in the shape of the source redshift distributions, such that the resulting IA parameter constraints are not a true reflection of the underlying IA  model.    This point is nicely illustrated in figure 5 of \citet{efstathiou/lemos:2018} where indirect constraints on the amplitude of the NLA model for the full galaxy sample are shown to vary from $ -6 < A_{\rm IA} < 6$ across a wide range of published cosmic shear surveys.  \citet{wright/etal:2020b} present another example, analysing mock galaxy catalogues to improve the accuracy of the priors on the redshift uncertainty for the previous KiDS data release.  In this case the fiducial IA constraint $A_{\rm IA} = 0.95 \pm 0.67$ reduces to $A_{\rm IA} = 0.28 \pm 0.59$ with the inclusion of a more accurate redshift uncertainty model.   As the intrinsic alignment model depends on the cosmology, any tendency for the model to incorrectly absorb redshift errors would inadvertently lead to unexpected biases in the cosmological parameter constraints.   The more freedom afforded to the IA model, the more opportunity there is for such biases to occur.    Given this concern, we choose to adopt the minimal IA model freedom afforded by current direct observational IA constraints.  In this way any unaccounted errors in our redshift distributions can be detected through a poor goodness-of-fit of the combined cosmological and IA model, as seen, for our second tomographic redshift bin, in the KiDS-1000 internal consistency analysis of \citet[][appendix B.2]{asgari/etal:2020}.   

Our choice of the one-parameter NLA model is motivated by the IA halo model analysis of \citet{fortuna/etal:2020}.   In this analysis central red galaxies are modelled using the  \citet{hirata/seljak:2004} model, with an amplitude $A_{\rm red}$ and a luminosity dependent scaling $\propto L^\beta$.   Combining the observed constraints from \citet{joachimi/etal:2011, singh/etal:2015} they model a simple power-law scaling model with $A_{\rm red} = 5.3 \pm 0.6$ and $\beta=1.2 \pm 0.4$.   The additional constraints from \citet{johnston/etal:2019} motivate a broken power-law model with $A_{\rm red} = 5.1 \pm 1.0$ and $\beta_{L \geq L_0}=1.2 \pm 0.4$, which they also consider.    Central blue galaxies follow the \citet{hirata/seljak:2004} model, with a \citet{johnston/etal:2019} constrained amplitude of $A_{\rm blue} = 0.2 \pm 0.4$ and no luminosity dependence, as there is no observational evidence to support this.    The small-scale satellite galaxy alignments are modelled following \citet{schneider/bridle:2010}.  The amplitude, radial and luminosity dependence of the alignment of red and blue satellite galaxies within their host halo is independently constrained following \citet{georgiou/etal:2019}.     The large-scale galaxy alignment signal is considered to be sourced solely by central galaxies, as there is currently no observational evidence to support large-scale (two-halo) alignments between the different satellite populations.

The direct IA observational constraints adopted by \citet{fortuna/etal:2020} are determined from low redshift galaxy samples, where the properties of the full galaxy population differ significantly from the high redshift galaxy population analysed in KiDS-1000.  Given that the fundamental physical processes that underpin the intrinsic alignment mechanisms are unlikely to significantly evolve out to $z \sim 1$, however, these galaxy-type-specific constraints are relevant at the redshifts sampled by KiDS, provided the evolution in the relative fractions, luminosities and distribution of the red and blue, central and satellite, galaxy populations are accurately modelled.   

\citet{fortuna/etal:2020} determine a sample of plausible IA models for a KiDS-like survey.  They combine the different galaxy-type IA contributions using a halo occupation distribution model taken from the MICE mock galaxy catalogues \citep{fosalba/etal:2015}, with a magnitude limit corresponding to KiDS-depth.   The range of full-population models encompasses the observational uncertainty on the IA model parameters for each galaxy type.    Analysing these IA halo-models with the NLA model, they find the highest NLA model amplitude of $A_{\rm IA}=0.44 \pm 0.13$ when adopting a broken power-law for the red-central luminosity scaling.    This is consistent with the \citet{asgari/etal:2020} KiDS-1000 COSEBIs cosmic shear constraints with $A_{\rm IA}=0.26 ^{+0.42}_{-0.34}$, the KiDS-1000 band power cosmic shear constraints with $A_{\rm IA}=0.97 ^{+0.29}_{-0.38}$ and the KiDS-1000 \tttp band power constraints with $A_{\rm IA}=1.07^{+0.27}_{-0.31}$, with the largest $1.9 \sigma$ difference found for the \tttp results.    

Adopting the NLA model in a cosmological parameter inference of a mock KiDS-like data vector of the cosmic shear signal contaminated by the range of different IA halo models allowed by observations, \citet{fortuna/etal:2020} conclude that the redshift dependence of the true IA halo model is not large enough to bias the cosmological parameters in a KiDS-like cosmic shear analysis with the NLA model.   \citet{asgari/etal:2020} nevertheless explore one extension of the NLA model, with the inclusion of a redshift-dependent scaling term, which does not change the recovered KiDS-1000 cosmic shear MAP $S_8$ value.  It only serves to increase the marginal credible region of $S_8$ by $\sim 10\%$.   Given these analyses, we choose to adopt a minimal one-parameter NLA model in our \tttp analysis, but recognise that in future surveys such as LSST and {\it Euclid}, this adhoc model will no longer be sufficient given the expected statistical power of these next-generation surveys \citep{blazek/etal:2019, fortuna/etal:2020}.

\section{Sensitivity tests}
\label{app:sensitivity}

\begin{figure*}
	\begin{center}
		\includegraphics[width=\textwidth]{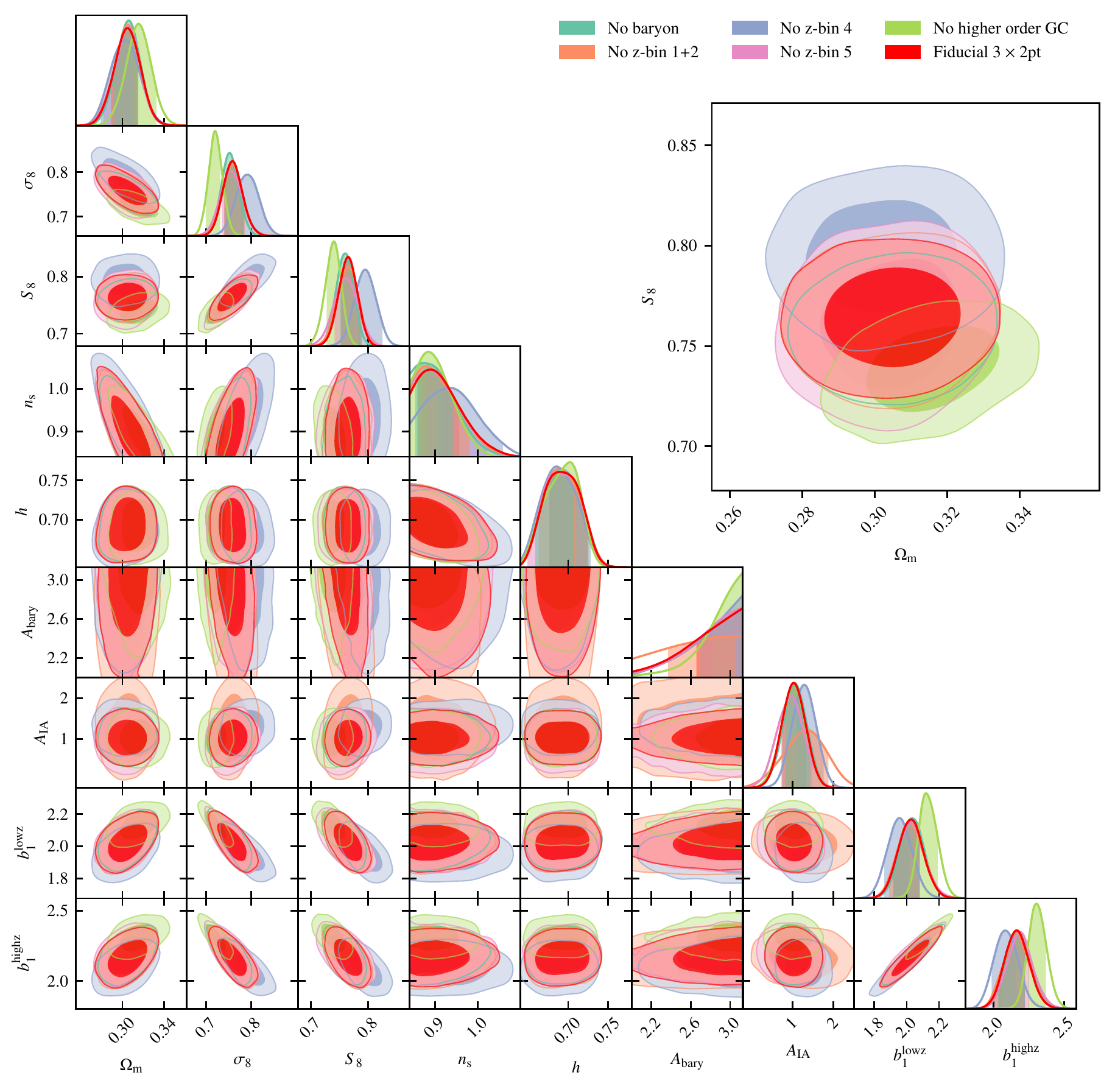}
		\caption{Marginalised posterior distributions for the extended set of cosmological parameters shown in Fig.~\ref{fig:cosmology-params-all}, comparing the fiducial \tttp analysis (red) to a selection of our sensitivity test analyses where we ignore the impact of baryon feedback (the `No baryon' case, sea-green), limit the analysis to a linear galaxy bias model (the `No higher order GC' case, lime-green), and remove individual tomographic bins from our weak lensing observables (orange, purple and pink).}
		\label{fig:sensitivity_tests}
	\end{center}
\end{figure*}

In this appendix we present, with Fig.~\ref{fig:sensitivity_tests}, the marginalised posterior distributions for a selection of the \tttp sensitivity tests explored in Sect.~\ref{sec:results}.  
These tests were all shown to recover consistent constraints on $S_8$, in Fig.~\ref{fig:S8comp}.  Here we explore these tests in more detail.    

We compare our analysis which fully marginalises over our uncertainty in the baryon feedback parameter (red), with our `No baryon' case (sea-green), where $A_{\rm bary}=3.13$, corresponding to the non-linear matter power spectrum for a dark-matter only cosmology.   
Here we find very little difference, as our \tttp analysis already favours high values of $A_{\rm bary}$.   
Our choice of scales and adoption of the band-power cosmic shear statistic for this analysis also makes us less sensitive to uncertainties in the baryon feedback parameter, compared to a standard two-point correlation function analysis \citep{asgari/etal:2020_KD}.

The removal of the two highest photometric redshift bins (blue and purple) primarily impacts $S_8$.   
These two bins carry the majority of the signal-to-noise in our analysis, and so it is not surprising that the removal of nearly half the constraining data, in each case, can result in $\sim\! 1\,\sigma$ changes in the recovered $S_8$ (see for example the discussion in Appendix~\ref{app:expectedoffsets}).   
We refer the reader to \citet{asgari/etal:inprep} where we present a detailed internal consistency analysis of the cosmic shear signal, following \citet{kohlinger/etal:2019}, \citep[see also][]{efstathiou/lemos:2018},  concluding that the two highest photometric redshift bins are consistent with the full data set.   
A potential $\sim\! 3\,\sigma$-level flag is, however, raised in \citet{asgari/etal:inprep} 
over the internal consistency of the second tomographic bin.  
In our analysis where we remove the two lowest photometric redshift bins (orange), we find that these bins contribute very little to the $S_8$ constraint and only serve to tighten the constraints on the intrinsic alignment parameter $A_{\rm IA}$.

Finally we turn to the galaxy bias test (lime green), where we limit the analysis to a linear galaxy bias model, $b_1$, setting all higher-order bias terms in Eq. (\ref{eq:pgg}) to zero\footnote{Removing all higher-order galaxy bias terms also requires the uncoupling of $b_1$ from $\gamma_2$.}, as well as imposing a Gaussian galaxy velocity distribution by setting $a_{\rm vir}$ to zero.   
We can see that this biases the recovered cosmological constraints, as the amplitude of the linear galaxy bias increases in an attempt to model the enhanced power on small scales that the non-linear bias induces.   
The erroneous increase in $b_1$ leads to a decrease in the recovered $S_8$, and also an overall reduction in the goodness-of-fit of the model with $\chi^2_{\rm MAP} = 379.8$ for $\sim\! 300$ degrees of freedom.  This result should serve as a point of caution for \tttp analyses that adopt an effective linear bias model \citep[see also the discussion in][]{asgari/etal:2020}, although we note that our analysis is particularly sensitive to the galaxy bias model given the high signal-to-noise BOSS clustering observations that probe physical scales as low as $s_{\rm min}= 20 h^{-1}\, {\rm Mpc}$.

\section{Expected $S_8$ differences between partially overlapping weak lensing surveys}
\label{app:expectedoffsets}
In this appendix we construct a simple model to estimate the expected statistical fluctuation in $S_8$ constraints from partially overlapping weak lensing surveys, specifically our previous \citep[KV450,][]{wright/etal:2020b}, and current KiDS analyses.   Assuming that we only wish to constrain the overall amplitude of the measurement, the inference can be approximated by a linear least-squares problem, where $S_{8}$, the parameter of interest, is Gaussian distributed. 

In the following, measurements derived from the KV450 footprint are denoted by an $X$ in the subscript, while measurements on the new area added for KiDS-1000 are denoted with a $Y$. 
Measurements derived from the full KiDS-1000 footprint are denoted with a $Z$. 
Now, let $S^{\rm KV450}_{X}$ denote the value of $S_{8}$ estimated from the KV450 footprint, using the KV450 methodology; $S^\text{KiDS-1000}_{Y}$ the value estimated from the newly added area using the KiDS-1000 methodology; and $S^\text{KiDS-1000}_{Z}$ the KiDS-1000 value estimated from the full footprint. 
Approximating the estimates from disjoint footprints as independent, the $S_8$ measurement from the full KiDS-1000 area is then given by the area-weighted average\footnote{The amplitude of the weak lensing signal scales roughly with $S_8$ \citep{jain/seljak:1997}, with an uncertainty variance that scales with the inverse survey area \citep{schneider/etal:2002}.}
\begin{equation}
  S^\text{KiDS-1000}_Z = \frac{A_X S^\text{KiDS-1000}_X + A_Y S^\text{KiDS-1000}_Y}{A_Z} \, .
\end{equation}
Here $A_X$ is the effective area of the KV450 footprint, $A_Z$ is the effective area of the KiDS-1000 footprint, and $A_Y = A_Z - A_X$, is the additional area added between the two data releases.  
The uncertainty $\sigma_{X}^\text{KiDS-1000}$ on the measurement $S^\text{KiDS-1000}_X$ is then related to the uncertainty $\sigma_Z^\text{KiDS-1000}$ on measurements $S^\text{KiDS-1000}_Z$, as 
\be
\sigma_{X}^\text{KiDS-1000} = \sqrt{\frac{A_Z}{A_X}}\, \sigma_Z^\text{KiDS-1000} \, ,
\ee
and analogously for $\sigma_{Y}^\text{KiDS-1000}$. 
It is important to note that the uncertainty $\sigma^{\rm KV450}_{X}$ differs from $\sigma_{X}^\text{KiDS-1000}$ due to differences in the KiDS-1000 and KV450 methodologies, such as the adopted $m$-calibration uncertainty. 

We define $\Delta = S^\text{KiDS-1000}_Z - S^{\rm KV450}_X$, the offset between the KiDS-1000 and KV450 $S_{8}$ measurements.  
The uncertainty on $\Delta$ is then given by
\be
\sigma_\Delta^2 = \left(\sigma_X^{\rm KV450}\right)^2 + \left(\sigma_Z^\text{KiDS-1000}\right)^2 - 2\sqrt{\frac{A_X}{A_Z}}\sigma_X^{\rm KV450}\sigma_Z^\text{KiDS-1000} \, ,
\ee
where we have assumed a 100\% correlation between the KiDS-1000 and KV450 measurements within the KV450 footprint area $X$, but approximate the areas $X$ and $Y$ as fully uncorrelated.  
This approximation neglects the large-scale correlations between $X$ and $Y$.  Given that the majority of the error budget for KiDS-1000 stems from the random shape noise component, which is independent between $X$ and $Y$, this approximation is sufficient for this toy model.
The absolute difference, $|\Delta|$ has the cumulative distribution function, CDF, 
\begin{equation}
   \mathrm{CDF}(x) = \mathrm{erf}\left(\frac{x}{\sigma_\Delta\sqrt{2}}\right) \, ,
\end{equation}
and an expectation value $\mathrm{E}[|\Delta|] = \sqrt{2/\pi} \,\sigma_\Delta$. 

We now compare the actual KV450 and KiDS-1000 $S_8$ constraints, given the effective areas $A_{\rm KV450} = A_X = 341.3\,\mathrm{deg}^{2}$ and $A_{\rm KiDS-1000} = A_Z = 777.4\,\mathrm{deg}^{2}$. 
Using the marginal $S_8$ constraints from the two-point shear correlation function analysis, comparing \citet[][KV450: $S_8 = 0.716^{+0.043}_{-0.038}$]{wright/etal:2020b} with \citet[][KiDS-1000: $S_8=0.768^{+0.016}_{-0.020}$]{asgari/etal:inprep}, 
we find $|\Delta|= 0.052 = 1.6\sigma_\Delta$, which can be compared against the expected offset of $\mathrm{E}[|\Delta|]=0.026$. 
We expect to find an offset of this, or a larger magnitude, 10\% of the time.  

This simple model analysis is sufficient to conclude that the increase in $S_8$ that we find between KV450 and KiDS-1000 is consistent with the expectation from simple statistical fluctuations.    A complete assessment could be conducted by analysing a reasonable fraction of the $20\,000$ KiDS mock catalogues from \citet{joachimi/etal:inprep}, with the KiDS-1000 and KV450 footprints imposed.  Unfortunately this is, however, out-of-scope for this analysis, given the significant compute power that would be incurred.

\section{Tension estimators}
\label{app:tensionest}
In this Appendix we review the range of tension estimators employed in Sect.~\ref{sec:planck_comp} to quantify the tension between the cosmological parameter constraints from {\it Planck} and our KiDS-1000-BOSS \tttp analysis.   These are in addition to the standard Gaussian offset measure, $\tau$, in Eq.~(\ref{eq: std_tension}), and the Bayes factor, $R$, in Eq.~(\ref{equ:bayes-factor}).

\subsection{Hellinger tension measure: $d_{\rm H}$}
\label{app:hellinger}
To assess the tension between the marginal $S_8$ distributions of two experiments, we have employed the widely used Gaussian-approximation measure, $\tau$, given in Eq.~(\ref{eq: std_tension}).  As an alternative that does not depend on the Gaussian assumption, we construct a sample-based, one-dimensional tension estimator adopting the Hellinger distance $d_{\rm H}$. The Hellinger distance is widely used in statistics as a stable metric in optimisation problems that require the comparison of distributions \citep[see for example.][]{beran77}. It is given by
\begin{equation}
\label{eq:hellinger_def}
d_{\rm H}^2 \bb{ p; q} = \frac{1}{2} \int {\rm d}x \bb{ \sqrt{p(x)} - \sqrt{q(x)} }^2\,,
\end{equation}
where the integral runs over the support of the PDFs of the distributions $p$ and $q$. We choose this distance definition as it is symmetric, avoids PDFs in the denominator and the associated instability for sample estimates, and has a closed-form expression for the case of two Gaussians:
\begin{equation}
\label{eq:hellinger_gauss}
d_{\rm H}^2 \bb{ {\cal N}(\mu_1,\sigma_1^2); {\cal N}(\mu_2,\sigma_2^2) } = 1 - \sqrt{ \frac{2 \sigma_1 \sigma_2}{\sigma_1^2 + \sigma_2^2} }\,  \exp \bc{ - \frac{(\mu_1 - \mu_2)^2}{4 (\sigma_1^2 + \sigma_2^2)} } , 
\end{equation}
where the normal distributions have means $\mu_{1,2}$, and variances $\sigma_{1,2}^2$. 

We estimate the Hellinger distance between our samples via a discretised version of Eq.~(\ref{eq:hellinger_def}). The PDFs are built from the samples in two ways: through binning into a histogram (with the number of bins chosen to scale with the square root of the number of samples), and through kernel density estimation (using a Gaussian kernel with the width chosen by Scott's rule). The former approach will in general over-estimate tension as tails are underpopulated due to the discrete sampling, whereas the latter approach under-estimates tension due to the smoothing of the PDFs by the kernel density estimator. Empirically, we find that a simple average of the two approaches yields good accuracy over the few-$\sigma$ tension range that we are interested in.

We invert Eq.~(\ref{eq:hellinger_gauss}) to determine the expected difference in the means, $|\mu_1 - \mu_2|$ if the two distributions were Gaussian, for a given value of $d_{\rm H}$. To do so, we use the standard deviations estimated from the samples. The resulting difference in the means is inserted into Eq.~(\ref{eq: std_tension}) to obtain a measure of tension in the familiar $\sigma$ units\footnote{Our Hellinger tension measure is validated by comparing to Eq.~(\ref{eq: std_tension}) for Gaussian random variates with the same number of realisations as the sizes of our posterior samples 
We find that systematic deviations in the Hellinger tension are at $0.1\, \sigma$ or less for tension up to $4\,\sigma$, with a standard deviation of $\pm 0.05\, \sigma$.}.   Comparing the \tttp and {\it Planck} $S_8$ distributions, we find a Hellinger distance $d_{\rm H}=0.95$, which corresponds to a tension of $3.1\,\sigma$.

\subsection{Distribution of parameter shifts: $p_{\rm S}$}
\label{app:paramshift} 
Following \citet[][see also \citealt{kohlinger/etal:2019}]{Raveri2020}, we consider the distribution $P(\Delta\vec\theta)$ of the parameter shifts $\Delta\vec\theta = \vec\theta_{\it Planck} - \vec\theta_{{\rm 3\times2pt+}{\it Planck}}$. 
The significance of any shift can then be assessed by calculating how much of $P(\Delta\vec\theta)$ is enclosed within the iso-probability surface at $P(0)$:
\be
\label{equ:ps}
	p_{\rm S} = \int_{P(\Delta\vec\theta)>P(0)} P(\Delta\vec\theta)\diff\Delta\vec\theta \, .
\ee
In other words, how much of the distribution of parameter shifts $P(\Delta\vec\theta)$ is more likely than the probability of no shifts. 
Accurately evaluating the integral in Eq.~\eqref{equ:ps} when $P(\Delta\vec\theta)$ is not centred on zero requires very large numbers of samples to sufficiently cover the tails of the distribution. 
For a single parameter, $S_{8}$, we find $p_{\rm S} = 0.9981$, with a PTE of $0.0019$, or $3.1\,\sigma$. 
This is in agreement with the other single-parameter tension estimators \eqref{eq: std_tension} and \eqref{eq:hellinger_def}. 
For the five shared parameters between our \tttp analysis and {\it Planck} we find lower significances, around $2\, \sigma$, albeit with very large uncertainties owing to the insufficient sampling of the tails of the parameter shift distribution. 

\subsection{Suspiciousness: $S$}
\label{app:sus} 
\citet{handley/lemos:2019} propose the `suspiciousness' statistic $S$ based on the Bayes factor, $R$, but hardened against prior dependences.
They define 
\be
\label{equ:suspiciousness}
 \ln S = \ln R - \ln I \,,
\ee
where $\ln R$ is the logarithm of the Bayes factor in Eq.~\eqref{equ:bayes-factor}:
\be
\label{eqn:logR}
	\ln R = \ln Z_{{\rm 3\times2pt+}{\it Planck}}- \ln Z_{\rm 3\times2pt} - \ln Z_{\it Planck} \,,
\ee 
with the evidence $Z_{i}$ given by
\be
	Z = \int \mathcal{L}\pi\, \diff\vec\theta \,,
\ee
the integral of the likelihood $\mathcal{L}$ and prior $\pi$ over the parameters $\vec\theta$. 
The information ratio $I$ in Eq.~\eqref{equ:suspiciousness} is defined as
\be
 \ln I = \mathcal{D}_{\rm 3\times2pt} + \mathcal{D}_{\it Planck}  - \mathcal{D}_{{\rm 3\times2pt+}{\it Planck}} \,,
\ee
with $\mathcal{D}_{i}$ being the Kullback-Leibler divergence between the posterior $P$ and prior for probe $i$:
\begin{splitequation}
	\mathcal{D} &= \int P \ln \frac{P}{\pi}\diff\vec\theta = \int P \ln\mathcal{L}\, \diff\vec\theta - \ln Z \\
	&= \langle\ln\mathcal{L}\rangle_{P} - \ln Z \,.
\end{splitequation}
The second equality follows from Bayes theorem: $P = \mathcal{L}\pi \,/\, Z$. 
Using this definition of $\mathcal{D}$ allows us to rephrase the suspiciousness solely in terms of the expectation values of the log-likelihoods:
\begin{splitequation}
	\ln S &= \langle\ln\mathcal{L}_{{\rm 3\times2pt+}{\it Planck}}\rangle_{P_{{\rm 3\times2pt+}{\it Planck}}} - \langle\ln\mathcal{L}_{\rm 3\times2pt}\rangle_{P_{\rm 3\times2pt}} \\
	&\quad- \langle\ln\mathcal{L}_{\it Planck}\rangle_{P_{\it Planck}} \,.
\end{splitequation}
We find $\ln S=-2.0\pm0.1$. 
For Gaussian posteriors, the quantity $d-2\ln S$ is distributed as $\chi^2_{d}$, where $d=d_{\rm 3\times2pt} + d_{\it Planck} - d_{{\rm 3\times2pt+}{\it Planck}}$ is the difference in the Bayesian model dimensionalities.
This allows us to assign a probability of the observed suspiciousness under the assumption that the two data sets are in concordance. 
We calculate $d$, following \citet{handley/lemos:2019}, from the variances of the log-likelihoods. 
Other estimates of the Bayesian model dimensionality, such as those introduced in \citet{Raveri2019} yield very similar results. 
We calculate $d=3.3\pm0.7$ and therefore conclude that the probability of observing our measured suspiciousness statistic is $0.08\pm0.02$, or $1.8\pm0.1\,\sigma$. 

\subsection{Goodness-of-fit change: $Q_{\rm DMAP}$}
\label{app:QDMAP} 
\citet{Raveri2019} introduce the $Q_{\rm DMAP}$ statistic that quantifies how the goodness-of-fit changes when two data sets are combined, based on the difference of the log-likelihoods, $\ln\mathcal{L}(\vec\theta^{\rm MAP}_{i})$, at the MAP of the respective posteriors\footnote{As an interesting aside, we note that the combined quantity $Q_{\rm DMAP} + 4\ln S$ corresponds to the twice the deviance information criterion ratio introduced in \citet{joudaki/etal:2017}.},
\be
\label{eqn:QDMAP}
	Q_{\rm DMAP} =  2\ln\mathcal{L}(\vec\theta^{\rm MAP}_{\rm 3\times2pt}) + 2\ln\mathcal{L}(\vec\theta^{\rm MAP}_{\it Planck}) - 2\ln\mathcal{L}(\vec\theta^{\rm MAP}_{{\rm 3\times2pt+}{\it Planck}}) \,.
\ee
For Gaussian posteriors, the sampling distribution of $Q_{\rm DMAP}$ is $\chi^{2}$-distributed with $d_{\rm 3\times2pt} + d_{\it Planck} - d_{{\rm 3\times2pt+}{\it Planck}}$ degrees of freedom. 
Here we follow \citet{Raveri2019} and calculate the model dimensionalities by $d_{i} = N - \mathrm{tr}[\mathcal{C}_{\pi}^{-1}\mathcal{C}_{P}]$, where $N$ is the total number of varied parameters, $\mathcal{C}_{\pi}$ is the prior parameter covariance, and $\mathcal{C}_{P}$ is the posterior parameter covariance. 
We find consistent results when using the variance of the log-likelihoods, the approach taken for the suspiciousness calculation.
We find $Q_{\rm DMAP} = 9.5$, with $d=3.6$, corresponding to a probability of 0.037, or $2.1\, \sigma$.

\subsection{Parameter difference in multiple dimensions: $Q_{\rm UDM}$} 
\label{app:QUDM} 
\citet{Raveri2019} introduce the $Q_{\rm UDM}$ statistic that quantifies the difference between cosmological parameters across multiple dimensions.   Here
\be
\label{equ:qudm}
	Q_{\rm UDM} = \Delta\overline{\vec\theta}^{T}(\mathcal{C}_{\it Planck} - \mathcal{C}_{{\rm 3\times2pt+}{\it Planck}})^{-1}\Delta\overline{\vec\theta} \,,
\ee
where $\Delta\overline{\vec\theta} = \bar{\vec\theta}_{\it Planck} - \overline{\vec\theta}_{{\rm 3\times2pt+}{\it Planck}}$ is the shift in the mean of the shared parameters, and $\mathcal{C}_{\it Planck}$ and $\mathcal{C}_{{\rm 3\times2pt+}{\it Planck}}$ are the posterior parameter covariances for {\it Planck} and \tttp+{\it Planck}, respectively. 
The estimator Eq.~\eqref{equ:qudm} is asymmetric in the first data set if the posteriors are non-Gaussian. 
In this case, the data set with the more Gaussian posterior should be used, which in our case is {\it Planck}. 
We use \software{tensiometer}\footnote{\url{https://github.com/mraveri/tensiometer}} to compute $Q_{\rm UDM}$ and its associated number of degrees of freedom. 
We find $Q_{\rm UDM} = 7.7$ with $d = 3$. The sampling distribution is approximated by $\chi^{2}_{d}$, such that the probability of the observed value of $Q_{\rm UDM} $ is $0.054$, or $1.9\,\sigma$.

\section{Redundancy, pipeline validation and software review}
\label{app:codereview}

In this appendix we briefly review the redundancy in the KiDS-1000 analysis, from pixels through to parameters.   Multi-band pixel processing and photometry in the optical has been carried out on the full KiDS-1000 data set using two independent pipelines, {\sc AstroWISE} and {\sc THELI} \citep{begeman/etal:2013, erben/etal:2013}.  This approach allowed us to resolve a range of different issues, primarily in the astrometric solution, for the handful of problem fields that resisted automated processing.  We adopt the {\sc AstroWISE} reduction for multi-band photometry and the {\sc THELI} $r$-band reduction for object detection and weak lensing shape measurement \citep{kuijken/etal:2019}.  We consider only one shape measurement technique, {\it lens}fit \citep{miller/etal:2013}, but this is calibrated using a series of different image simulations that vary the input galaxy properties to assess the sensitivity of the shear estimator to: the presence of blending with unresolved and undetected objects, blending due to enhanced galaxy clustering, photometric redshift selection bias, size-ellipticity correlations in the galaxy properties, varying stellar density, and the choice of smooth or realistic galaxy profiles \citep[see][for details]{kannawadi/etal:2019, giblin/etal:inprep}.  

Our fiducial photometric redshift calibration has been compared to two additional independent calibration approaches, and has been validated on mock photometry catalogues \citep[see][for details]{wright/etal:2020, vandenbusch/etal:2020, hildebrandt/etal:inprep}.   Using a series of null-tests we have validated the resulting shear-redshift catalogues in \citet{giblin/etal:inprep}.    Our catalogue-to-observables pipeline, based on the two-point correlation function code {\sc TreeCorr} \citep{treecorr}, has been validated through an independent analysis using the alternative {\sc athena} package \citep{athena}, and through mock catalogue analysis \citep{joachimi/etal:inprep}.   Our cosmological inference code {\sc KCAP} with {\sc CAMB} \citep{lewis/etal:2000}, has been validated against the Core Cosmology Library \citep[CCL,][]{chisari/etal:2019} and through the recovery of input parameters in mock data analysis \citep{joachimi/etal:inprep}.  In \citet{asgari/etal:inprep} we also verify that we produce the same results in our cosmic-shear only analysis using a completely independent inference code based on {\sc MontePython} with {\sc CLASS} \citep{class, montepython, kohlinger/etal:2019,hildebrandt/etal:2020}.   Our analysis of BOSS follows \citet{sanchez/etal:2017}, which is in good agreement with the numerous independent parallel analyses of the same DR12 data set presented in \citet{alam/etal:2017}.

In regards to software review, for the majority of cases we have adopted the `four-eye' approach, collaboratively building code through a git repository, with major updates reviewed through pull requests.   This approach has applied throughout the full pixels-to-parameters process, and to all types of software, both our significant tools, analysis code and, for the most part, simple paper support scripts.  Our repositories are open source\footnote{The software used to carry out the various analyses presented in this paper is open source at \href{https://github.com/KiDS-WL/Cat_to_Obs_K1000_P1}{github.com/KiDS-WL/Cat\textunderscore to\textunderscore Obs\textunderscore K1000\textunderscore P1} and \href{https://github.com/KiDS-WL/KCAP}{github.com/KiDS-WL/KCAP}. } for others to use and build upon, but with a caveat for users to recognise that we are not software engineers.   Adopting this style of software review for KiDS-1000 has been an extremely beneficial exercise for the team, with many lessons learnt for how to improve our software engineering skills and our approach to open source collaborative coding for future projects.     

\section{Post-unblinding analyses}
\label{app:unblinding}
This analysis was carried out `blind',  such that our final key result, our constraint on $S_8$, was unknown until all analysis choices were fixed.   Our blinding strategy creates three versions of the catalogue, where one is the truth, and the other two are modified to introduce up to a $\sim\! \pm 2\,\sigma$ deviation in the recovered value of $S_8$ \citep{kuijken/etal:2015, giblin/etal:inprep}.   In contrast to previous KiDS analyses, we set ourselves a challenge to only run our blind data analysis inference once, after fully developing and road-testing the pipeline using mock catalogues and data vectors \citep{joachimi/etal:inprep}.     Unexpectedly, the only analysis where we failed to meet this challenge were for data vectors that included galaxy clustering.   Here a bug in a naming convention in an updated {\sc CosmoSIS-CAMB} interface resulted in $\sim\! 0.5\,\sigma$ errors in the recovery of $\sigma_8$ in our initial BOSS re-analysis.  This was corrected and updated for our fiducial analysis before unblinding.

Our fiducial results, presented in Fig.~\ref{fig:cosmology-params} and tabulated in Appendix~\ref{app:parameter-constraints}, were carried out for all three blinds using a covariance matrix derived assuming a fiducial cosmology given by the best-fit parameters from \citet{troester/etal:2020}.  We reserved our iterated-covariance analysis, however, for post-unblinding, re-analysing the true data vector with an updated covariance matrix derived adopting the best-fit parameters from our initial blinded run of the true catalogue.   This iterative step changed our value for the full \tttp analysis $S_8$ by $0.2\,\sigma$ for the MAP constraint, and by $0.1\, \sigma$ for the marginal constraint.   Our tension-consistency analysis with the CMB constraints from Planck was also conducted after unblinding, as we were unblind to Planck throughout the process.

As we are only interested in relative shifts in the sensitivity tests presented in Fig~\ref{fig:sensitivity_tests}, these were only conducted for a single blind, and re-calculated post-unblinding for the true catalogue.  Wishing to be fully transparent in this appendix, we note that owing to limited resources the sensitivity test was not updated during the blinded phase of the project in order to correct for the {\sc CosmoSIS-CAMB} interface error which only impacted the BOSS constraints.   We argue that this was appropriate given the nature of the test and given that this error has been corrected post-unblinding.    Our cosmic shear and galaxy clustering analysis, featured in our internal consistency test in Fig.~\ref{fig:S8comp}, was also not re-processed blind after the BOSS error was corrected, as both our forecast in \citet{joachimi/etal:inprep}, and our initial analysis, confirmed that the constraints from our cosmic shear and galaxy clustering analysis were almost identical to constraints from the full \tttp analysis, which was correctly analysed for all three blinds.

As reported in \citet{giblin/etal:inprep}, co-author Kannawadi was unblinded early in the process to permit accurate calibration of the shear measurements in his re-analysis of KiDS-1000-like image simulations.   Co-authors Wright and Heymans also wish to record that they independently had a suspicion over which blind was the truth based on the changes that the SOM gold selection made to the blinded KiDS-1000 catalogue effective number density and ellipticity dispersion, compared to the impact the SOM selection made to the KiDS-VIKING-450 catalogue analysed in \citet{wright/etal:2020b}.  These suspicions were never shared with the rest of the team, nor discussed with each other, and on unblinding were found to both be false, and different!   This issue nevertheless highlighted to us the challenges of catalogue-level blinding for successive data releases where little has changed in the core data reduction process.  This is particularly relevant if the people working at the coal-face of catalogue production are the same people working on the cosmological analysis.   Future KiDS blinding will therefore likely adopt the approach advocated by \citet{sellentin:2020}, where the covariance matrix is modified in the analysis.

\end{appendix}


\end{document}